\documentclass[fleqn,usenatbib]{mnras}

\usepackage{newtxtext,newtxmath}

\usepackage[T1]{fontenc}

\DeclareRobustCommand{\VAN}[3]{#2}
\let\VANthebibliography\thebibliography
\def\thebibliography{\DeclareRobustCommand{\VAN}[3]{##3}\VANthebibliography}

\usepackage{graphicx}	
\usepackage{amsmath}	
\let\oldAA\AA
\renewcommand{\AA}{\text{\normalfont\oldAA}}

\title[A low frequency sub-arcsec view of 3C\,34 and 3C\,320]{A low frequency sub-arcsecond view of powerful radio galaxies in rich-cluster environments: 3C\,34 and 3C\,320}

\author[V. H. Mahatma]{
V.\,H.\,Mahatma,$^{1}$\thanks{E-mail: v.mahatma93@gmail.com}
A.\,Basu,$^{1}$
M.\,J.\,Hardcastle,$^{2}$
L.\,K.\,Morabito,$^{3,4}$
R.\,J.\,van Weeren$^{5}$
\\
$^{1}$Th\"uringer Landessternwarte, Sternwarte 5, 07778 Tautenburg, Germany\\
$^{2}$Centre for Astrophysics Research, Department of Physics, Astronomy and Mathematics, University of Hertfordshire, College Lane, Hatfield AL10 9AB, UK\\
$^{3}$Centre for Extragalactic Astronomy, Department of Physics, Durham University, Durham DH1 3LE, UK\\
$^{4}$Institute for Computational Cosmology, Department of Physics, University of Durham, South Road, Durham DH1 3LE, UK\\
$^{5}$Leiden Observatory, Leiden University, PO Box 9513, 2300 RA Leiden, The Netherlands
}

\date{Accepted 2023 February 01. Received 2023 February 01; in original form 2022 December 17}

\pubyear{2023}

\begin{document}
\label{firstpage}
\pagerange{\pageref{firstpage}--\pageref{lastpage}}
\maketitle

\begin{abstract}
Models of radio galaxy physics have been primarily based on high frequency ($\geqslant$1\,GHz) observations of their jets, hotspots, and lobes. Without highly resolved low frequency observations, which provide information on older plasma, our understanding of the dynamics of radio galaxies and their interaction with their environment is limited. Here, we present the first sub-arcsecond (0.3”) resolution images at 144\,MHz of two powerful radio galaxies situated in rich cluster environments, namely 3C\,34 and 3C\,320, using the International Low Frequency Array Telescope. We detect for the first time at low frequencies a plethora of structures in these objects, including strikingly large filaments across the base of the lobes in both sources, which are spatially associated with dense regions in the ambient medium. For 3C\,34 we report a spectral flattening in the region of the central filament, suggesting that the origin of the filaments is related to the presence of large-scale ordered magnetic fields. We also report periodic total intensity and spectral index banding of diffuse emission in the eastern lobe, seen for the first time in radio galaxy lobes. The hotspot complexes are resolved into multiple fragments of varying structure and spectral index; we discuss the implications for particle acceleration and jet termination models. We find at most smooth gradients in the spectral behaviour of the hotspot structure suggesting that particle acceleration, if present, may be occurring throughout the complex, in contrast to simple models, but different jet termination models may apply to both sources. 
\end{abstract}




\section{Introduction}
Deep radio observations of the jets, hotspots, and lobes of radio galaxies (or radio-loud AGN) over the past few decades have given us crucial insights into their energetics and dynamics. Particularly with the advent of sensitive instruments capable of observing nearby radio galaxies at sub-arcsecond resolution, primarily at frequencies greater than 1 GHz (e.g. Karl G. Jansky Very Large Array and e-MERLIN), it has been found that the synchrotron plasma in radio galaxy lobes and hotspots exhibit complex structures that challenge early models of radio galaxy evolution \citep[e.g.][]{fern97,hard97,leah97}. 

The `twin-exhaust' or `beam' models refined by \cite{sche74} and \cite{blan74} gave important qualitative understanding on radio galaxy energetics and dynamics: energy is transported from the central nucleus to hotspots by bipolar jets. The jet-head \textit{must} decelerate (near the hotspots) to conserve momentum, since the ambient medium is denser than the jet itself. This causes material to flow backward from the jet-head, forming the radio lobes which are, at the very least, in pressure balance with the surrounding medium as they continue to expand. If the expansion of the lobes is supersonic, they will drive a shock into the ambient medium. These particular studies considered simplistic uniform environments, but further (semi-) analytic models \citep[e.g.][]{turn15,hard18} refined this picture to consider radio source environments, such as power-law \citep{kais97}, $\beta$ model \citep[][]{beta} and halo mass dependent atmospheres \citep[][]{arna10}. Numerical simulations which include more detailed physics such as bulk relativistic flow, magnetic fields and spectral ageing have been performed for many years \citep[e.g.][]{turn15,engl16,yate22}, however the description of straight jets, two hotspots either side of the central engine and a spheroidal shell containing the radio source and its surrounding shocked gas that is symmetrical about the jet axis have remained principally unchanged.

Observationally, however, radio galaxies exhibit components that are far from this simplistic picture. Aside from the fact that Fanaroff-Riley type-I radio galaxies \citep[FR-I;][]{FR74} are morphologically different (since they do not form bright hotspots at the leading edge of the source), Fanaroff-Riley type II \citep[FR-II;][]{FR74} radio galaxies can include, but are not limited to: non-visible jets on at least one side of the source, compact knots along the visible jet, jet bending, multiple compact features that may all satisfy the defining criteria of a hotspot \citep[e.g.][]{leah97}, radio lobes that are non-spheroidal, `wings' and tails that tend to develop at the base of the lobes and expand transversely away from the jet axis (so-called X and Z-shaped radio galaxies; e.g. \citealt{cott20,mull21,gopa22}), filamentation \citep[e.g.][]{rama20} and fan structures of very low surface brightness adjacent to the lobes \citep[e.g. 3C\,351, 3C\,263;][]{brid94}. Deep Very Large Array (VLA) observations at sub-arcsecond resolution of the brightest radio galaxies (e.g. identified by the Third Cambridge Catalogue of radio sources, 3CR; \citealt{lain83}) have detected these complex morphological features at GHz frequencies for many years \citep[e.g.][]{rudn90,brid94,fern97,hard97,gilb04,mull06,shul19,rudn22}, but no empirical relationships between these features and the overall physical properties of radio galaxies (e.g. their environments) that can be adopted by dynamical models have been found. 

While more refined models are currently needed to better describe the physics of jets, lobes and hotspots, we are still missing information on the detailed spectral structure of radio galaxies down to low ($\sim100$\,MHz) frequencies. The aforementioned studies have been performed primarily at frequencies above 1\,GHz, where sub-arcsecond imaging (corresponding to a resolution element of $\sim0.5$\,kpc at $z=0.1$) was possible with past instruments. Since higher energy synchrotron-emitting electrons radiate quicker than lower energy electrons, the oldest plasma in radio galaxy lobes is detected only at the lowest frequencies (above the turnover frequency due to self-absorption or above a low-frequency energy cut-off). Spectral information over a large frequency range (between 100\,MHz and 8\,GHz, for example) at high resolution therefore allows us to distinguish between \textit{both} old and freshly accelerated particle populations in the lobes, for which we are missing understanding of at kpc-scale resolution. Standard models predict that the oldest plasma collects at the base of the lobes near the radio core, however recent studies discovering the development of large-scale filaments \citep[e.g.][]{rama20}, wings \citep[e.g.][]{cott20} and re-acceleration of the oldest plasma \citep[e.g.][]{dega17} has uncovered further questions on their evolution, and on their ubiquity amongst the population. Low frequencies also provide the required short baselines that probe structure on large angular scales, important for viewing the large-scale lobe-environment interactions seen in recent low frequency studies \citep[e.g.][]{hard19_ngc326}.

The interpretation of filamentation and other structures also lacks robustness without information on the ambient medium, to which the dynamical evolution is strongly linked. Recent low frequency (144-675\,MHz) observations of the radio galaxy NGC\,507 revealed an S-shaped filament emanating from the radio lobes, suggested as being introduced due to gas sloshing in its rich group environment \citep[][]{brie22}. Filaments associated with radio galaxies are becoming more frequently detected as instruments become more sensitive, and their formation is suggested to be driven by complex environmental interactions \citep[e.g.][]{hard19_ngc326,rudn21,know21,fana21,riseley22}, but further studies are needed to assert whether their formation is ubiquitous in the radio galaxy population in rich environments. Such interactions are also important for cluster science, as one of the unanswered fundamental questions concerns the origin of the large-scale diffuse radio emission (radio haloes and relics) found in an increasing number of deep observations of clusters \citep[see review by][]{vanw19}. The most plausible explanation is the enrichment of the medium with relativistic particles and magnetic fields from the lobes of radio galaxies. It is therefore important to understand how radio galaxy plasma couples to the cluster environment, the precise mechanisms of which are unknown. A low frequency study (combined with higher frequencies) of radio galaxies in cluster environments, with resolution that is sufficient to view the substructures of hotspots and lobes, is required to address these questions. 

The hotspot `structure' (or hotspot `complex'; \citealt{leah97}) in radio galaxy lobes is also important to understand. Many radio galaxies, when viewed at a physical resolution of the order 1 kpc, display multiple compact, bright and flat-spectrum features at the leading edge of a given lobe that meet the definition of a hotspot \citep[e.g.][]{lain82}. Many questions arise from this fact: the classical models of FR-II jet evolution assume jets terminate at a single hotspot where particle acceleration of the jet material occurs \citep[][]{blan74}. If there are multiple hotspots in radio galaxy lobes, are they all associated with the beam termination, and how do they form, evolve and contribute to the electron energy distribution of the radio source as a whole? There is typically a variation in brightness, compactness and spectral index between the multiple `hotspots' \citep{brid94}, while there is evidence to suggest particle acceleration is taking place in more than one hotspot in a given lobe \citep[][]{hard07}. Although sites of particle acceleration can unambiguously be identified using X-ray data that are well-described by synchrotron radiation, low frequency radio data (of the order 100\,MHz) can identify areas of flat-spectrum emission relative to steep-spectrum emission which may give clues to acceleration and the flow of plasma between hotspots.  

In this paper, the first of a series of papers describing sub-arcsecond resolution radio observations of the brightest radio galaxies at 144\,MHz using the International Low Frequency Array \citep[hereafter LOFAR;][]{vanh13}, we report observations of two rich cluster powerful radio galaxies: 3C\,34 and 3C\,320. Our aims are to describe and understand:
\begin{itemize}
    \item Their morphology seen at 144\,MHz at 0.3 arcsec resolution with respect to higher frequency data;
    \item The nature, evolution, and fate of the oldest plasma in the sources and their interaction with their cluster environments;
    \item The dynamics and origins of multiple hotspots and their association to jet termination models.
\end{itemize}
In section \ref{sect:data} we describe the target selection and data reduction procedures. In section \ref{sect:analysis} we describe our analysis of 3C\,34 and 3C\,320, using LOFAR 144\,MHz data, higher frequency radio data and other multi-wavelength data. We discuss our results in section \ref{sect:discussion} and summarize our conclusions in section \ref{sect:conclusions}. Throughout the paper, we define the spectral index $\alpha$ in the sense $S\propto\nu^{-\alpha}$. We use a $\Lambda$CDM cosmology in which $H_0=67.4$ km s$^{-1}$ Mpc$^{-1}$, $\Omega_M=0.315$ and $\Omega_{\Lambda}=0.685$ \citep{planck20}.

\section{Data}
\label{sect:data}
\subsection{Target selection}
With sensitivity and image fidelity as a primary driver, LOFAR is the most suitable array for a deep low frequency study –- the high band antennas (HBA), observing at a typical central frequency of 144\,MHz, are distributed such that it offers excellent $uv$-coverage at the shortest and longest baselines (with a largest angular scale of $\sim$70 arcsec after station combination, discussed in Section \ref{sect:datareduction}). The LOFAR Two Metre Sky Survey (LoTSS; \citealt{shim19,shim22}), an ongoing project to survey the entire northern sky at 6\,arcsec resolution, processing only the Dutch array, achieves a median RMS of $\sim83\,\rm \upmu Jy\,beam^{-1}$. Use of the international stations, which are included in observations of some fields and provide baselines up to 2000 km, can typically achieve sub-arcsecond ($\sim$0.3) resolution at a similar sensitivity \citep[][]{mora22}, and is ideal for our science. We used the LoTSS\footnote{\url{https://lofar-surveys.org/dr2_release.html}} Data Release 2 (LoTSS DR2; \citealt{shim22}) to select 3CRR \citep[][]{lain83} sources -- these are the brightest ($\geqslant10$\,Jy at 178\,MHz) radio galaxies for which there are a wealth of available ancillary data and have been studied for many years \citep[e.g.][]{brid94,leah97,hard97} and offer complementary information to the LoTSS data. From the 3CRR sources observed by LoTSS at the time of selection\footnote{The LoTSS survey was $\sim30$\% complete at the time of selection}, we selected sources having a suitable nearby delay calibrator. We also needed to ensure the target is within a 1.25 degree radius of the LoTSS pointing primary beam -- the typical field of view (FOV) of the HBA international stations has a 2.5 degree diameter. Unfortunately, a number of LoTSS observations had been performed without the international stations, which limited our selection. With a handful of 3C objects remaining, we selected 3C\,34 and 3C\,320 as two of the brightest FR-II radio galaxies residing in cluster environments, sufficiently large in angular size (> 50 beam widths across their linear extent at 0.3 arcsec), showing evidence of multiple hotspots, and having available higher frequency radio data at the same resolution and multi-wavelength data describing their surrounding environment.

\subsection{3C\,34}
3C\,34 is an extended FR-II radio galaxy with a 178-MHz luminosity of $L_{178}=2.7\times10^{27}\,\rm W\,Hz^{-1}$ \citep[][]{lain83}, at a redshift of $z=0.69$ \citep[][]{spin85}. It lies at the centre of a rich compact cluster, with extended emission line gas from the host galaxy  overlapping the radio lobes \citep[][]{mcca95}. An [O{\sc iii}] emission line image was reported where the emission is elongated along the jet axis, a commonly found phenomenon for ($z>0.6$) powerful radio galaxies known as the alignment effect, suggested as describing jet-induced star-formation \citep{mcca87}. Optical \textit{Hubble Space Telescope} (HST) images show that the host has diffuse morphology, possibly as a result of the superposition of cluster-member galaxies \citep[][]{mcca95}. Deeper HST imaging revealed the presence of a long and narrow region of blue optical emission in the middle of the western lobe, suggested to also describe jet-induced star-formation in a cluster member \citep[][]{best97}, and supported by observations of stronger Faraday depolarization (between 1\,GHz and 5\,GHz) in this region \citep[][]{john95}. In the radio, both lobes are known to have a double-hotspot structure, and a jet is claimed to have been detected on both sides of the source (a jet is clearly detected on the western lobe while jet knots appear on the eastern lobe; \citealt{john95}), and is therefore likely to be orientated close to the plane of the sky. The back-flowing material in the lobes also have evidence for a deflection at the base to form `wings' \citep[][]{john95}. The consistent distribution of electric field vector angles suggest that these wings are part of a large scale back-flow originating from the hotspots, and in general they display high degrees of polarization \citep[][]{john95}. 3C\,34 is therefore an exemplary source to study jet-induced star-formation, large-scale radio wings, multiple hotspot structures and jet knots. 

\subsection{3C\,320}
3C320 is a classical FR-II radio galaxy with a radio luminosity of $L_{178}=3.3\times10^{27}\,\rm W\,Hz^{-1}$, at a redshift of $z=0.342$. It is identified as a brightest cluster galaxy (BCG) in the centre of a rich X-ray-emitting cluster \citep[][]{massaro13}, where a large-scale shock surrounding the lobes was interpreted from the X-ray emission. This shock was analysed with deep 0.5-5\,keV \textit{Chandra} observations \citep[][]{maha20}, while \cite{vags19} studied the large X-ray cavities associated with the radio lobes. While the shocks are clear evidence of AGN feedback, \cite{maha20} determined the jet power ($Q_{\rm jet}=1.05\times10^{38}$\,W) to be consistent with the cooling luminosity of X-ray photons in the cluster ($L_{\rm cool}=8.48\times10^{37}$\,W), suggesting that the radio source is sufficiently powerful to overcome radiative losses in the intracluster medium (ICM). There is also a central bar of X-ray emitting material between the lobes or cavities, indicating a dense region of accreting gas. A complex hotspot structure in the eastern lobe was also detected with the 1.5\,GHz e-MERLIN observations presented by \cite{maha20}, while the western lobe shows a clear jet with knots in their deep 5\,GHz VLA data. The X-ray emission of the ICM shows an elongated spheroidal morphology in projection, perhaps tracing a relatively recent merger. Due to its relatively small angular size of 18\,arcsec, past studies at radio frequencies are severely limited for this source. 

\subsection{Data processing}
\label{sect:datareduction}
\subsubsection{LOFAR}
3C\,34 and 3C\,320 were observed in LoTSS pointings P018+31 and P233+35 under project codes LC6\_015 and LC10\_010 on the dates of 2017 January 16 and 2018 September 14, respectively. Both pointings were observed for a total of 8 hours. We extracted the target data from the Long Term Archive\footnote{\url{https://lta.lofar.eu/Lofar}} (LTA), as well as the 10-minute observations of the primary flux calibrators (3C\,196 and 3C\,295 for 3C\,34 and 3C\,320, respectively) for these pointings, observed just before the target pointings. 231 sub-bands were extracted as individual measurement sets for all four data sets (two targets and two flux calibrators) each with 16\,channels and a 1\,second integration time, in the frequency range 120.3\,MHz to 167.7\,MHz (sub-bands at a higher frequency are prone to considerable RFI), with a central frequency of 144\,MHz. Both 3C\,196 and 3C\,295 are resolved for LOFAR with the international array, so we used high resolution models fixed to the \cite{scai12} flux density scale to ensure good solutions. For the pointing containing 3C\,34 a total of 74 LOFAR stations were used (the Dutch array consists of 24 core stations that are each divided into two sub-stations and 14 remote stations), including 12 international stations (non-Dutch). The pointing containing 3C\,320 was observed with a similar number of stations, but with the addition of a station in Ireland, giving the Ireland-Poland baseline as the longest baseline available to LoTSS observations (a $uv$ distance of $\sim2\times10^6\lambda$).   

First, we ran \texttt{PREFACTOR}\footnote{\url{https://github.com/lmorabit/prefactor}} \citep{dega19} on the calibrator data which first flags (using \texttt{AOFLAGGER}; \citealt{offr12}) and averages the data, then sets the flux density scale and corrects the calibrator data for the polarization alignment (between $XX$ and $YY$), bandpass, clock and total electron content (TEC) for all stations. The bandpasses for the international stations were typically a factor of 3 higher in amplitude than the Dutch stations, reflecting the higher station gains as more tiles are used. These corrections are then applied to the Dutch stations, and a phase-only calibration is performed on those stations using the TGSS-ADR1 sky model \citep{inte17}. The rotation measure for all the stations is also found, to help correct for bulk changes in the TEC. The pre-determined corrections (clock, bandpass, rotation measure) are then applied to the target.

From the data, which have at this point been only corrected for direction-independent effects, we used the foundational calibration strategy described by \cite{mora22} to calibrate the international stations. We created two data sets (using \texttt{DPPP}; \citealt{vand18}) which are phase-shifted to the most suitable\footnote{Most compact, bright and nearest to the target.} nearby calibrators\footnote{0.64 degrees and 0.49 degrees away from 3C\,34 and 3C\,320, respectively} from the Long Baseline Calibrator Survey \citep[LBCS;][]{jack22} and to the target. In these data sets the core stations are combined into a single super station to provide a sensitive reference antenna (\citealt{mora22}), and the data are averaged in frequency and time, leading to a final frequency resolution of 48\,kHz per channel and a time resolution of 8\,sec. We derived phase and amplitude calibration solutions for the delay calibrator following the method and pipeline described by \cite{vanw21}. These solutions were merged into a single file in the Hierarchical data format version 5 \citep[][]{hdf5} and applied to the target data. Since these solutions are direction-dependent, the final calibration procedure involved performing total electron content (TEC) and phase corrections on the target, in an iterative self-calibration procedure until convergence. In this step for the first self-calibration iteration we used a sky model for the target, using a Gaussian model formed by applying \texttt{pybdsf} to the 5\,GHz 0.3 arcsecond VLA images of 3C\,34 and 3C\,320 (see full details below), to correct the astrometry and help convergence. 

\begin{figure*}
    \centering
    \includegraphics[scale=0.45,trim={0.2cm 0.2cm 0.2cm 0.2cm},clip]{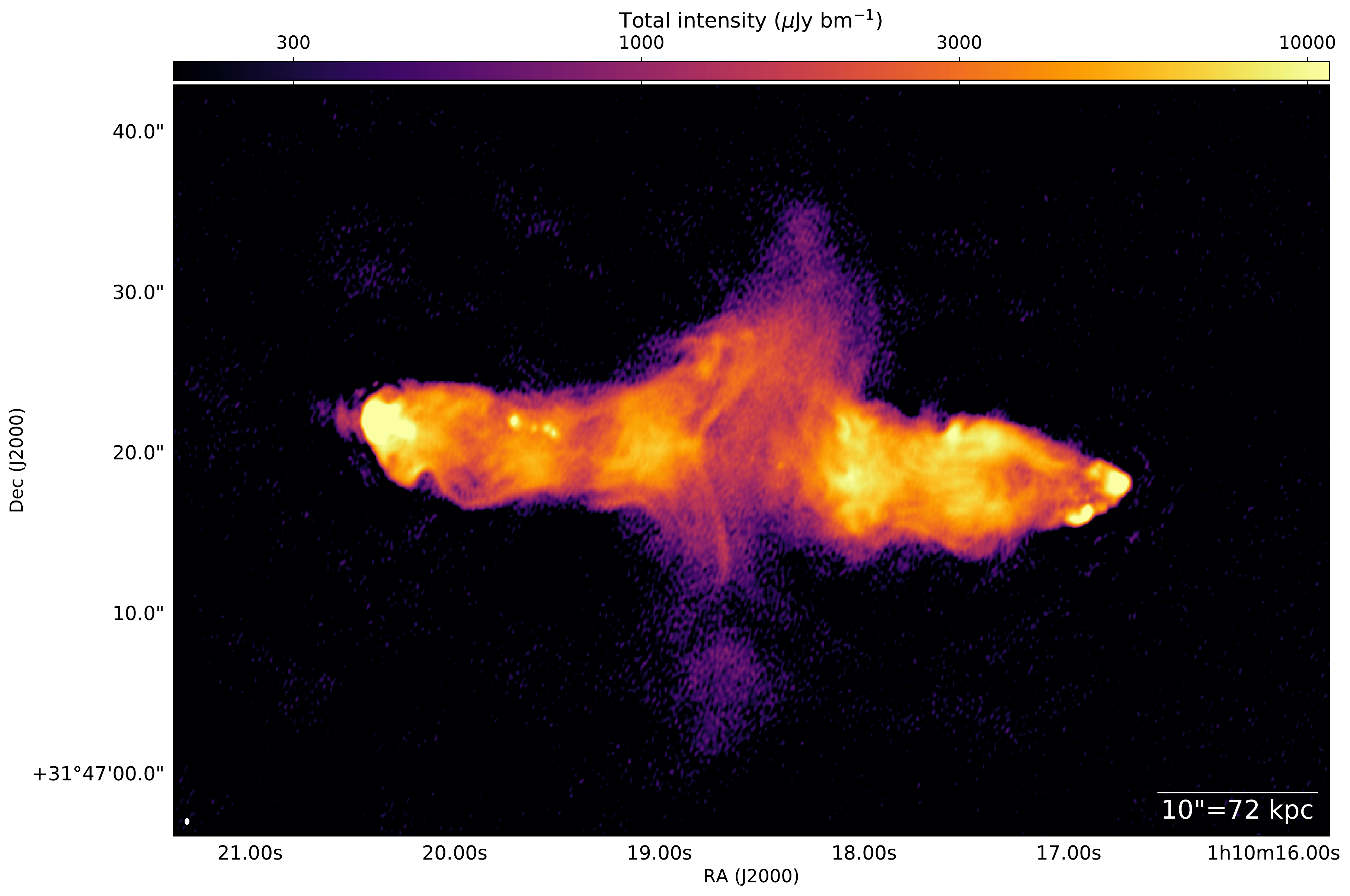}\\
    \includegraphics[scale=0.45,trim={0.2cm 0.2cm 0.2cm 0.2cm},clip]{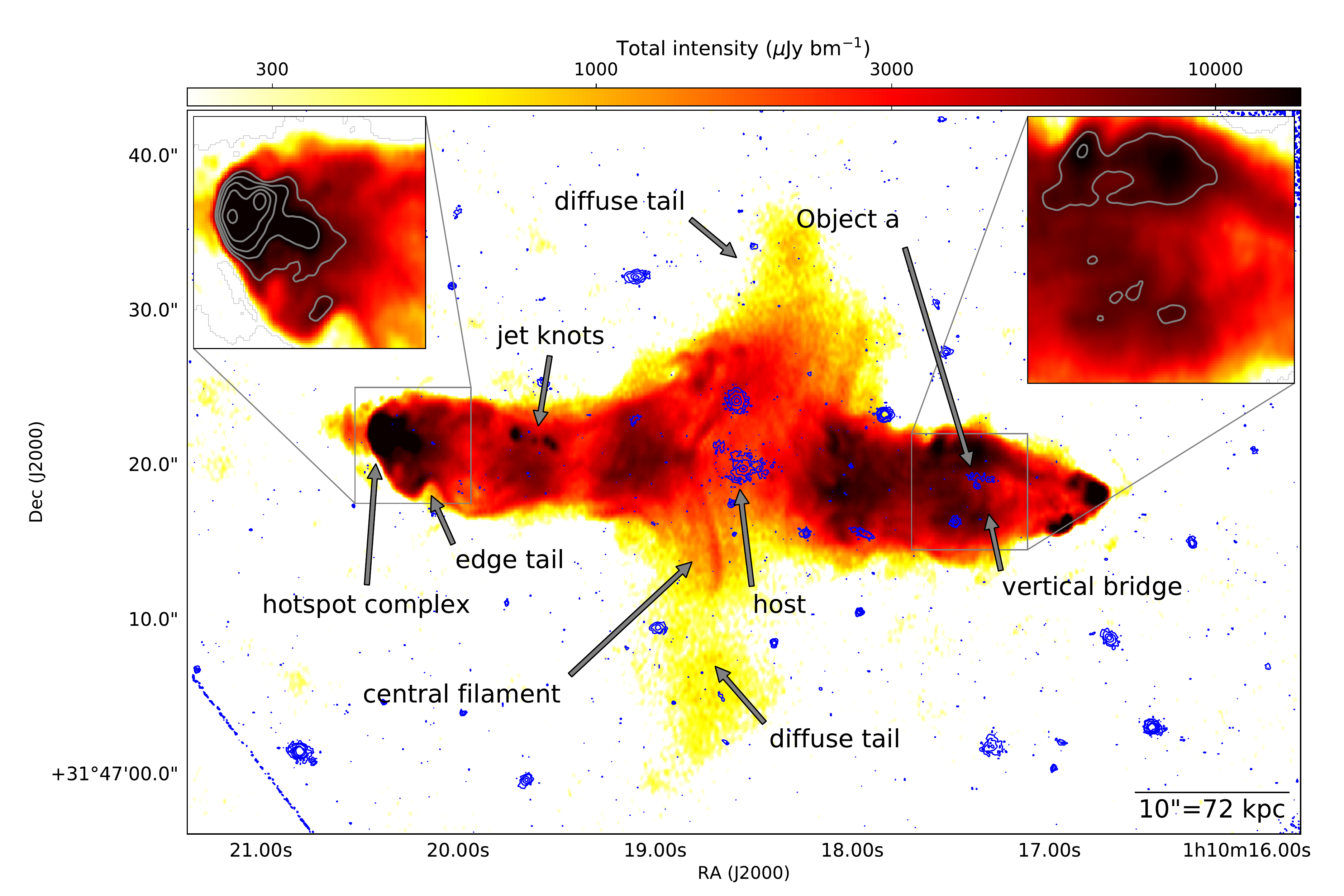}
    \caption{Top: LOFAR 144\,MHz image of 3C\,34 with a synthesized beam of 0.24$\times0.40$ arcsec and a background RMS of 109\,$\mu$Jy beam$^{-1}$. Bottom: LOFAR 144\,MHz image of 3C\,34 with optical HST contours (blue) with the f785LP filter ($\lambda=8620\AA$). Insets show zoom-ins of the eastern hotspot complex and of the limb brightening either size of a vertical bridge crossing `Object a', a region of jet-induced star-formation \citep[][]{best97}. The zoom-ins are overlaid with LOFAR contours (grey) from the same data, describing $50\sigma$ multiplied by 2 to the power of integers from 1 to 10, chosen to highlight compact structures. Other interesting features are annotated and discussed in the text. The LOFAR image beam size is shown in the top panel.} 
    \label{fig:3c34}
\end{figure*}

\begin{figure*}
    \centering
    \includegraphics[scale=0.45,trim={0.2cm 0.2cm 0.2cm 0.2cm},clip]{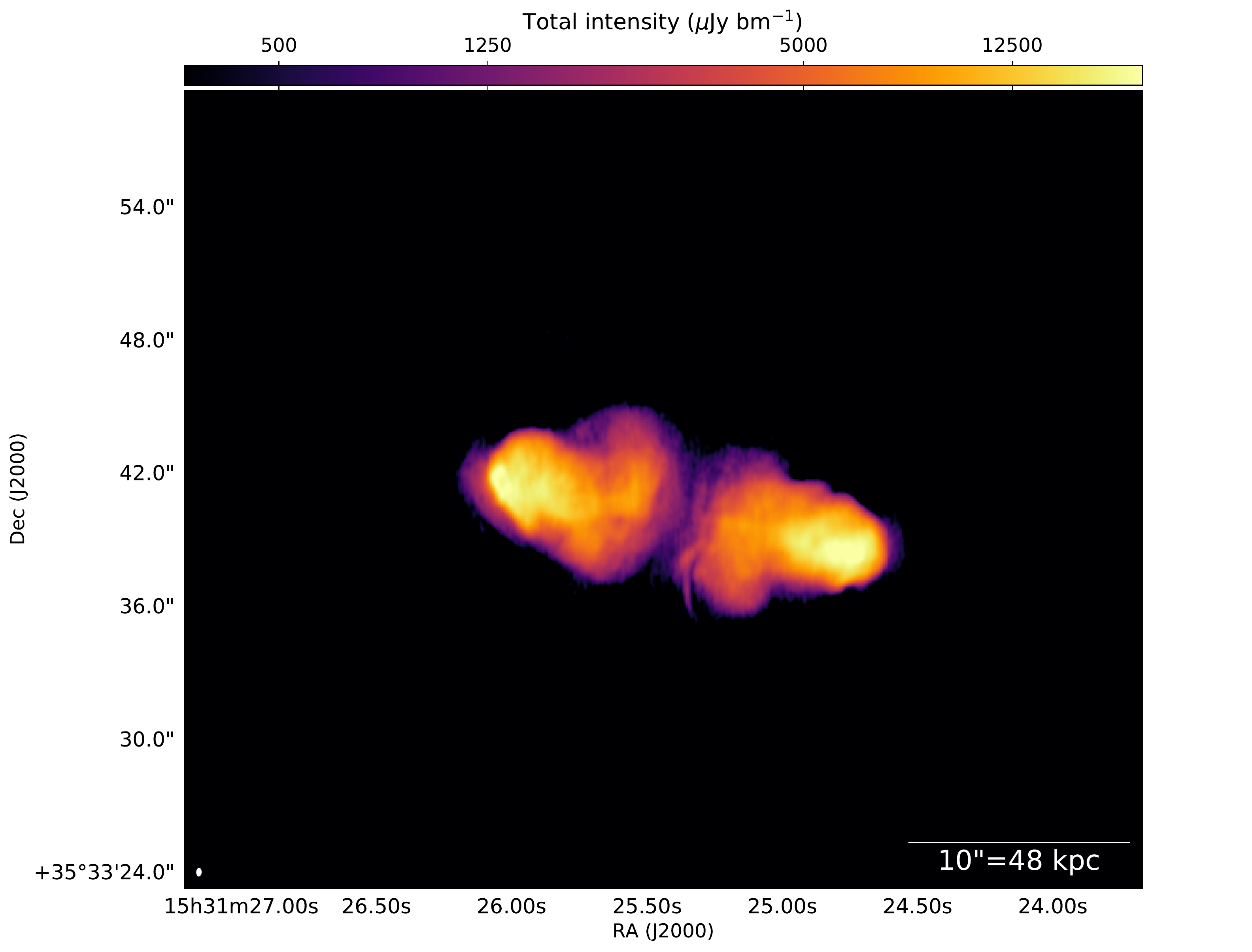}\\ 
    \hspace{0.4cm}
    \includegraphics[scale=0.45,trim={0.2cm 0.2cm 0.2cm 0.2cm},clip]{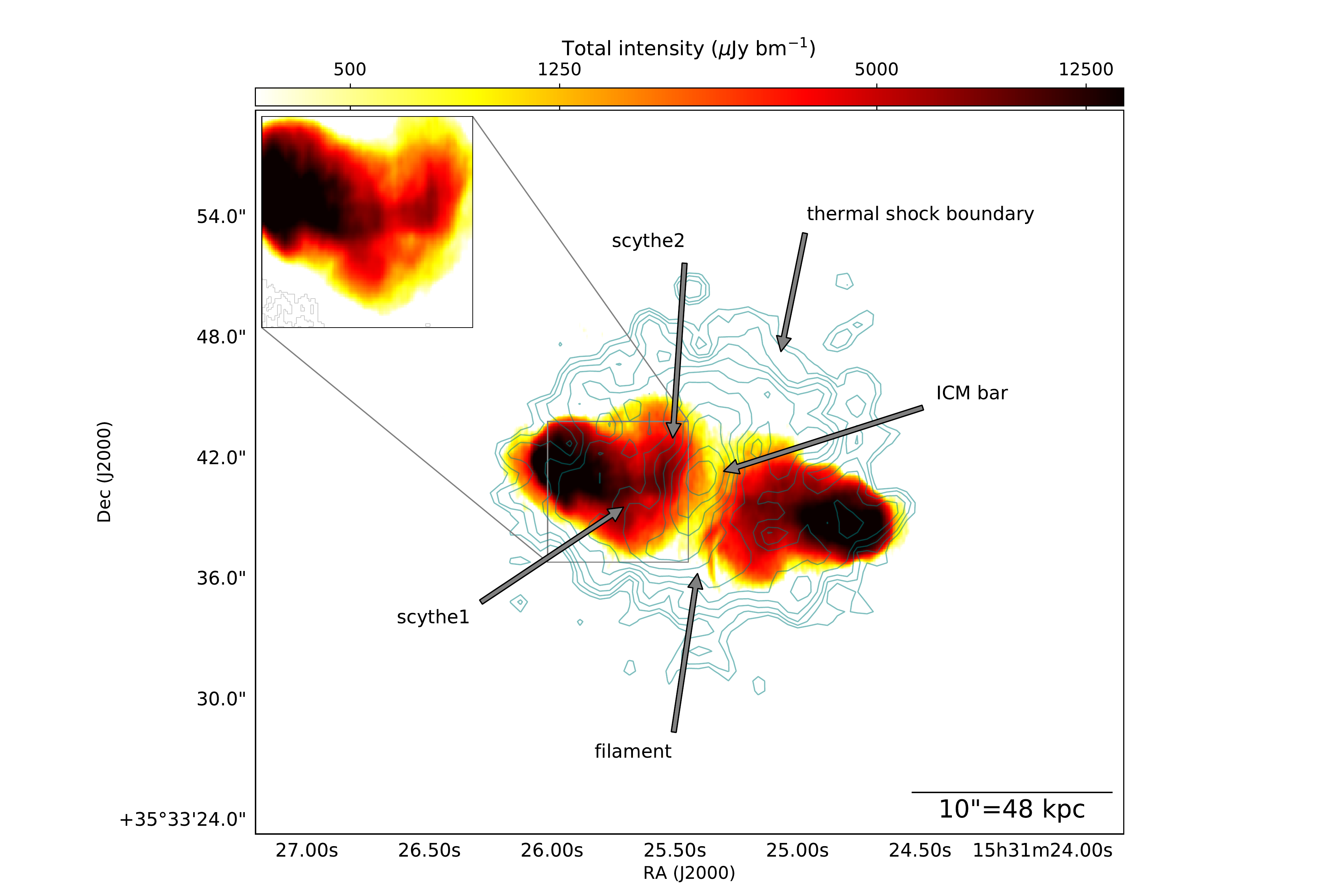}
    \caption{Top: LOFAR 144\,MHz image of 3C\,320 with the international array with a synthesized beam of 0.33$\times0.18$ arcsec and a background RMS of 77 $\mu$Jy beam$^{-1}$. Bottom: \textit{Chandra} 0.5-5 keV X-ray contours of the hot gas ICM surrounding 3C\,320 (teal), overlaid on the LOFAR 144 MHz image. Contours are at 3$\sigma$ plus $\sigma$ to the power of integers from 1 to 20. The scythes (see inset), filament and the thermal shock boundary driven by 3C\,320 are shown with arrows.}
    \label{fig:3c320}
\end{figure*}

The final full-bandwidth images were produced with \texttt{WSCLEAN} \citep[][]{offr14} on the corrected data sets, using multiscale \citep[][]{offr17} and multi-frequency \texttt{CLEAN}ing, and are shown on the top panels of Figures\,\ref{fig:3c34} and \ref{fig:3c320}. The background RMS noises in these images are $109\,\rm \upmu Jy\,beam^{-1}$ and $77\,\rm \upmu Jy\,beam^{-1}$, for 3C\,34 and 3C\,320, respectively. We achieve dynamic ranges of 1047 for 3C\,34 (Figure \ref{fig:3c34}) and 391 for 3C\,320 (Figure \ref{fig:3c320}). 

\begin{figure*}
    \centering
    \includegraphics[scale=0.45,trim={0.2cm 0.2cm 0.2cm 0.2cm},clip]{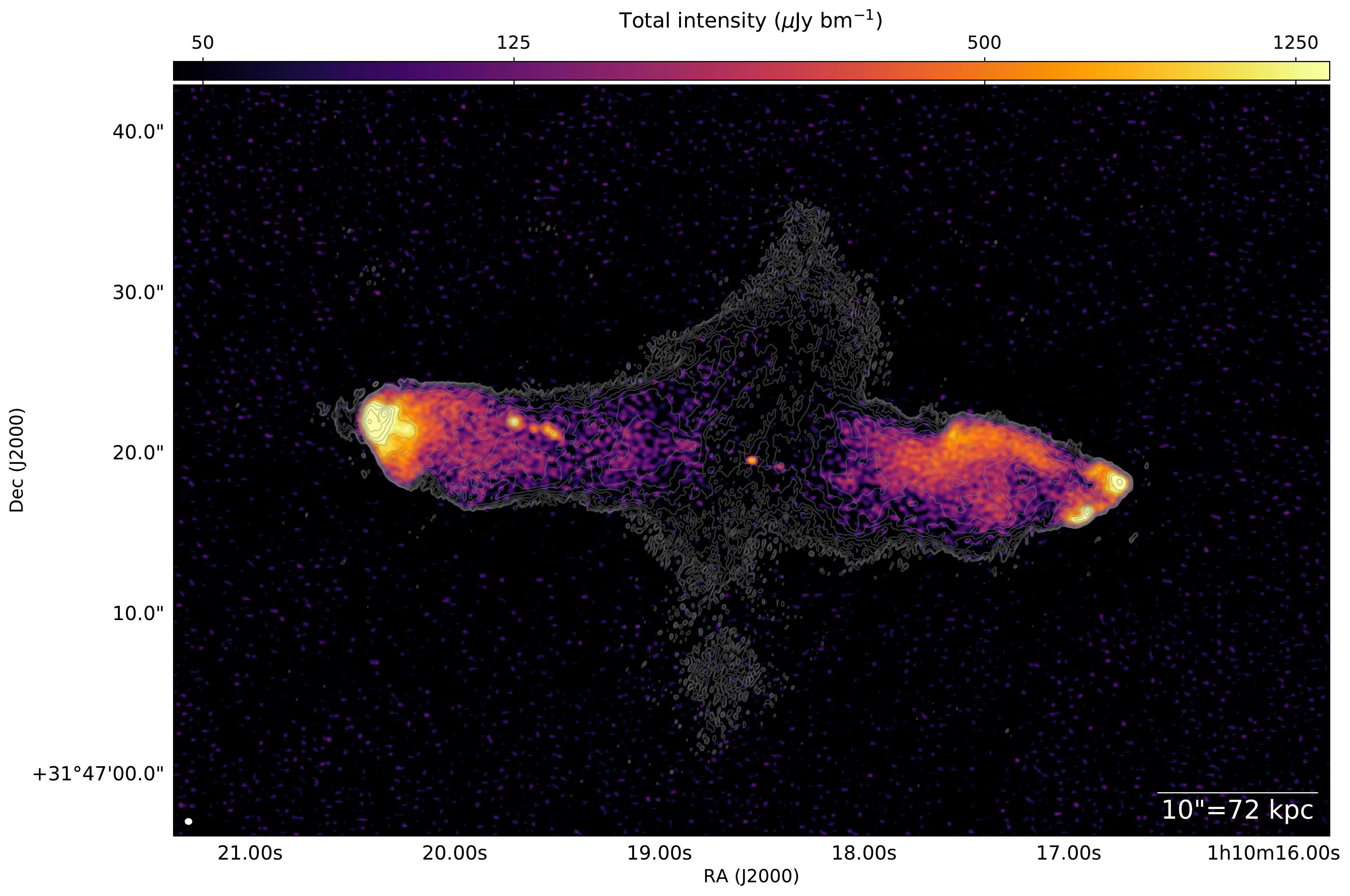}
    \caption{3C\,34 VLA 5000 MHz image using the combined archival A-, AnB- and C-array data, with a synthesized beam of 0.39$\times 0.34$ arcsec and a background RMS of 40$\mu$\,Jy\,beam$^{-1}$. Contours are of the 144 MHz image, describing 3$\sigma$ plus $\sigma$ multiplied by 1.5 to the power of integers from 1 to 20. The synthesized beam shape of the VLA data are shown on the lower left.}
    \label{fig:3c34_cband}
\end{figure*}
\begin{figure*}
    \centering
    \includegraphics[scale=0.45,trim={0.2cm 0.2cm 0.2cm 0.2cm},clip]{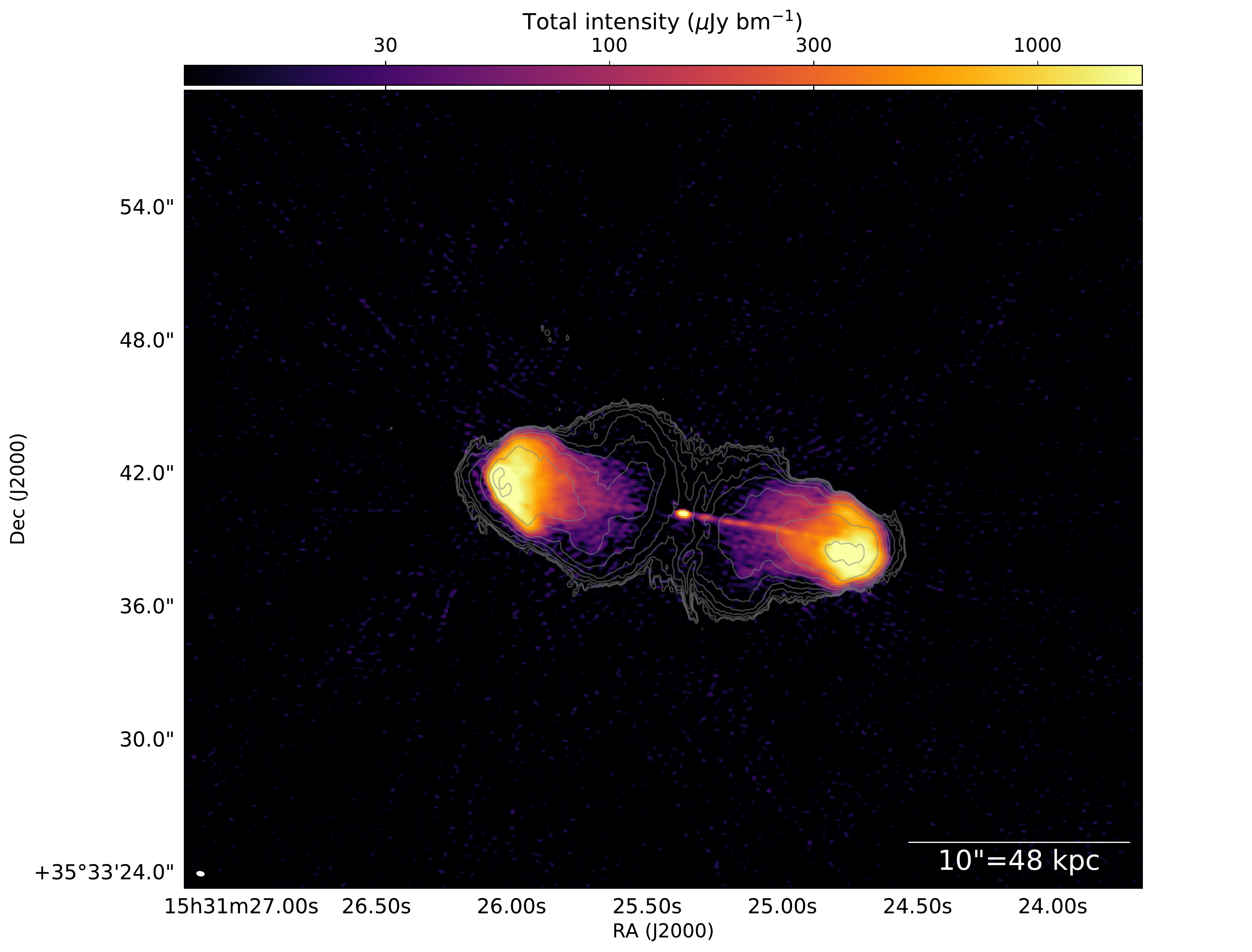}
    \caption{3C\,320 VLA 6 GHz image with the A and B array with a synthesized beam of 0.32$\times$0.18 arcsec and a background RMS of 6\,$\mu$Jy\,beam$^{-1}$, with LOFAR 144 MHz contours overlaid. Contours are of the 144 MHz image, describing 3$\sigma$ plus $\sigma$ multiplied by 1.5 to the power of integers from 1 to 20. Beam shapes are shown on the lower left.}
    \label{fig:3c320_cband}
\end{figure*}

\subsubsection{VLA and e-MERLIN}
We use archival 1.5\,GHz and 5\,GHz data to complement the LOFAR data in this study. To approximately match the dense $uv$-coverage obtained by the LOFAR observations in the shortest spacings, we combine multi-array data with the VLA. Due to the smaller angular size of 3C320, we additionally use e-MERLIN data in combination with the VLA at L band to provide $\lesssim0.3$ arcsec resolution. 

For 3C\,34, we use A, AnB and C-array data at 5\,GHz, and A-array data at 1.5\,GHz. Table~\ref{tab:vla} lists the observations we extracted from the NRAO archive. We reduced the data sets in the standard manner using \texttt{AIPS}. 3C\,286 was observed as the primary flux calibrator for the 1.5\,GHz data and for the A and AnB array data at 5\,GHz, and 3C\,48 for the 5\,GHz C-array data, applying the \cite{baars77} flux scale in the process. This was used to set the flux density of the phase calibrators (0133+476 for the 1.5\,GHz data and the 5\,GHz A and AnB array data). BL Lac was used as a phase calibrator for the 5\,GHz C-array data, allowing us to calibrate the complex gains and transferring these to 3C\,34. For the 5\,GHz C-array data, we followed the original analysis of these data from \cite{john95} by allowing for an increase of 3.7\% of the flux density of 3C\,48 since the \cite{baars77} measurement at the time of the observations. No bandpass or polarization calibration was performed. The data set was subsequently flagged for RFI using the \texttt{CASA} \citep{mcmu07} task \texttt{rflag} after calibration. The calibrated data for the target 3C34 were split and self-calibrated using \texttt{CASA}, with several rounds of phase calibration until the solution interval equated integration time (10\,s). Self-calibration for the 5\,GHz AnB and C-array datasets were initiated using a model using the self-calibrated A-array data, to ensure astrometric accuracy between the datasets. The final images were produced using the \texttt{CLEAN} algorithm on all four data sets. The 5\,GHz combined A, AnB and C array image is shown in Figure \ref{fig:3c34_cband}, with LOFAR contours overlaid, with a total flux density of 0.40 Jy, consistent with that of \cite{john95}.

For 3C\,320, being only 18\,arcsec in angular size, we use VLA A-array and B-array data at 6\,GHz, and VLA A-array data and e-MERLIN data at 1.5\,GHz (calibration, combination and imaging were discussed and presented by \citealt{maha20}, and we refer the reader to their study). The 6\,GHz VLA image is shown in Figure~\ref{fig:3c320_cband} with LOFAR 144\,MHz contours overlaid.

\begin{table*}
    \centering
    \begin{tabular}{c|c|c|c|c|c|c|c|c}
    \hline
    Source & Project & Array & Frequency (MHz) & Date & Exposure (min) & Beam (arcsec) & Flux density (Jy) & RMS ($\mu$Jy beam$^{-1}$)\\ \hline
    3C34 & LC6\_015 & LOFAR & 120-168 & 2017 Jan 16 & 480 & 0.24$\times$0.40 & 18.76 & 109 \\
        & AG247 & VLA-A & 1417-1662$^{\dagger}$ & 1987 Sept 11 & 556 & 1.21$\times$0.97 & 1.48 & 139 \\
        & AF213 & VLA-A & 4820-4870 & 1991 Jul 17 & 163 & 0.43$\times$0.38 & 0.26 & 40\\
        & AP380 & VLA-AnB & 4820-4870 & 1991 Dec 22 & 80 & 1.40$\times$1.11 & 0.40 & 42 \\
        & AG247 & VLA-C & 4715-4965$^{*}$ & 1988 Mar 29 & 47 & 6.87$\times$5.14 & 0.40 & 125 \\ \hline
    3C320 & LC10\_010 & LOFAR & 120-168 & 2018 Sept 14 & 480 & 0.33$\times$0.18 & 9.78 & 77 \\
        & 15A-420 & VLA-A & 1010-1970 & 2015 Aug 1 & 240 & 1.53$\times$0.87 & 1.81 & 67 \\
        & CY4223 & e-MERLIN & 1250-1700 & 2018 Mar 22 & 960 & 0.16$\times$0.15 & 1.02 & 30\\
        & 15A-420 & VLA-A & 3980-7900 & 2015 Aug 1 & 240 & 0.42$\times$0.22 & 0.40 & 7\\
        & 15A-420 & VLA-B & 3980-7900 & 2015 Feb 15 & 90 & 1.12$\times$0.74 & 0.43 & 35\\
        \hline
    \end{tabular}
    \caption{Radio observations of 3C\,34 and 3C\,320 used in this study. $^{\dagger}$Simultaneous sub-bands at 1417\,MHz and 1662\,MHz were observed, each with a bandwidth of 12.5\,MHz. $^{*}$Simultaneous sub-bands at 4715\,MHz and 4965\,MHz were observed, each with a bandwidth of 50\,MHz. Note that beam sizes are dependent on the weighting scheme used during imaging, with Briggs robust values between -1 and 0.5.}
    \label{tab:vla}
\end{table*}
\subsubsection{Spectral data and image alignment}
Here we state the methods used to obtain resolved spectral index maps for 3C\,34 and 3C\,320, using the higher frequency radio data discussed in Section \ref{sect:data}. Due to the lack of sub-arcsecond data at 1400\,MHz, we produce a sub-arcsecond spectral index map only using 144\,MHz and 5000\,MHz data for 3C\,34, giving two-point spectral indices at a resolution of 0.38 arcsecond to study the hotspots. For the diffuse emission in 3C\,34, we produce spectral index maps at 1.2 arcsecond without the 5000\,MHz data due to the low signal-to-noise for the steep-spectrum regions around the base of the lobes. For 3C\,320, given the high resolution e-MERLIN data at 1400\,MHz, we produce 0.2 arcsecond resolution maps across the frequency coverage (144, 1400\,MHz, 6000\,MHz), giving four-point spectral indices by splitting the 6000\,MHz data. To mitigate the effects of different calibration strategies introducing (sub)-pixel shifts in the astrometry, all the maps were aligned by Gaussian-fitting of point-like sources, in our case the hotspots, detected in each map \citep{harw19}. After aligning the maps, we used \texttt{BRATS} \citep[][]{harwood13} to produce spectral index maps, using a 5$\sigma$ total intensity detection threshold at each frequency.

As a caution, we note the uncertainties on the integrated flux densities that affect our resolved spectral index maps as a systematic offset. The data sets presented here are calibrated using different flux density scales -- the LOFAR data are on the \cite{scai12} scale, the VLA data for 3C\,34 are on the \cite{baars77} scale and the VLA data for 3C\,320 are on the \cite{perl17} scale. To estimate the magnitude of the flux density scale uncertainty introduced by this, we compared the flux density scales for the primary flux calibrators used in the data presented. For 3C\,34, we compared the \cite{baars77} scale to the \cite{scai12} scale at 144\,MHz for 3C\,286, the flux calibrator for the VLA observations at 1.5\,GHz and at 5\,GHz, finding a difference of 18.6\% (higher in the \citealt{baars77} scale). For 3C\,320, we compared the \cite{perl17} scale to the \cite{scai12} scale at 144\,MHz for 3C\,286, also the flux calibrator for the VLA higher frequency observations, finding a difference of 13.12\% (higher in the \citealt{perl17} scale). 

To cross-check the systematic flux density uncertainties on the LOFAR data, we compared the measured source flux density with that measured in the 6C survey \citep[][]{hale88,hale93} at 151\,MHz (calibrated using measurements of Cygnus A), scaled to 144\,MHz assuming a spectral index of 0.5. With these comparisons, 3C\,34 has an uncertainty of 18\% (higher in the LOFAR data), while 3C\,320 has an integrated flux density uncertainty of 16\% (higher in the LOFAR data). These factors, while large, are likely dominated by absolute flux density scale errors mentioned above, but also by imperfect beam models (which depend on a given LOFAR pointing) during the processing, meaning that the relative gains between the flux density calibrator and the target are not well determined. For the LoTSS DR2 catalogue using only the Dutch array, \cite{shim22} implement a procedure to align the flux density scale with 6C and NVSS, leading to an overall flux density uncertainty of less than 10\%, with a further potential error of 10\% due to positional variations in the LoTSS flux density scale, which also depend on position from the pointing centre \citep[][]{shim22}. Given that these uncertainties are on the Dutch array only, and that the addition of the international stations resulting in stronger ionospheric effects act to exacerbate flux density errors, we suggest that these uncertainties are not extreme, given the data. Flux density uncertainties in the range 10-20\% have been used for spectral analysis in recent studies of 3C sources using the full international array of LOFAR \citep[][]{swei22,kapp22}. It is important to note that these integrated flux density errors are systematic, and they do not apply to differences between points in highly resolved spectral index maps, particularly with large gaps in frequency coverage in the data presented. Nevertheless, and unless otherwise stated, we adopt these errors (combined with measurement errors based on background RMS) for the purposes of spectral index fitting, with the caveat that all quoted values are uncertain by a systematic offset predominantly. For the VLA L-band data for both 3C\,34 and 3C\,320, we adopt uncertainties of 10\% and 14\%, respectively, based on measurements using the Green Bank Telescope (GBT) at 1400 MHz, catalogued by \cite{kell69}, which themselves have flux density uncertainties of 15\% and are therefore consistent with our measurements. For the VLA C-band data for 3C\,34, we adopt an uncertainty of 5\% based on measurements using the GBT at 4830\,MHz \citep{grif90}. For the VLA C-band data for 3C\,320, we use the standard flux density calibration uncertainty of $\sim2$\% \citep[][]{perl17}. We analyse the resulting spectral index maps for both sources in Section \ref{sect:particle_acceleration}.

\section{Analysis}
\label{sect:analysis}
\begin{table}
    \centering
    \begin{tabular}{c|c|c|c}
    \hline
        Object & $z$ & LAS (") & $L_{178}$ (W Hz$^{-1}$) \\\hline
        3C34 &  0.691 & 49 & 2.7$\times10^{27}$\\
        3C320 & 0.342 & 20 & 3.3$\times10^{27}$\\
    \hline
    \end{tabular}
    \caption{Redshift, largest angular size and 178 MHz radio luminosity measured in the 3CRR survey.}
    \label{tab:my_label}
\end{table}

\subsection{Radio morphology}

\subsubsection{3C34}
In Figure~\ref{fig:3c34} (top panel) we present the LOFAR 144 MHz image at $\sim$0.3 arcsec resolution, and with optical HST contours\footnote{HST image courtesy of Phillip Best} overlaid (bottom panel). A plethora of compact and diffuse emission are observed in the 144\,MHz data, but the most striking feature is the large-scale filament across the base of the lobes spanning $\sim$16 arcsec in projected linear size (corresponding to $\sim$114\,kpc at the distance of 3C\,34), seen in this source for the first time. \cite{john95} discovered apparent wings at the base of the lobes, which are generally described as large-scale backflows that expand transversely from the jet direction at the base of the lobes in powerful sources, but no filament was seen with the 1.2 arcsec angular resolution of their data at 1500\,MHz, nor with the 0.3 arcsec angular resolution of their data at 5000\,MHz. At 144\,MHz at $\sim0.3$\,arcsec resolution (see Figure \ref{fig:3c34_cband} for a comparison with 5000 MHz), this emission is resolved into a central filament and diffuse tails, as annotated in the bottom panel of Figure \ref{fig:3c34}. The diffuse tails, extended in the North-South direction, could be an example of old plasma being distributed into the ICM by the central filament. The region of the central filament has been reported to have a high degree ($\gtrsim50$\% in some places) of polarization \citep[][]{john95} at > 1\,GHz, indicating ordered magnetic fields which could belong to the central filament. Unfortunately, techniques for robust polarization calibration for LOFAR did not exist at the time of writing, limiting our study to the total intensity at 144\,MHz. 

Filamentary structures are seen throughout the source -- the southern edge of the eastern lobe shows tails which seem to follow a trajectory starting from  just south of the leading hotspot to the base of the lobe (marked as `edge tail' in the bottom panel). The western lobe also shows such structures throughout. The leading western hotspot is connected with a faint tail (above `Object a') in the northern edge of the western lobe, which extends further downstream towards the core. There is also a very faint small-scale tail (marked as `vertical bridge') that bridges this region and another high-surface brightness region towards the south, while crossing `Object a' (see Section \ref{sect:environment}). It is clear that these filaments are produced in regions of highest turbulence (the lobes have higher depolarization ratios with distance from the hotspots; \citealt{john95}), but we discuss their nature through environmental and spectral analysis in the proceeding sections. 

\subsubsection{3C320}
In Figure \ref{fig:3c320} (top panel) we present the 144 MHz LOFAR image of 3C320, at a resolution of $\sim0.3$ arcsec. The source is dominated by diffuse emission at 144\,MHz, and the lobe material is seen to extend transversely to the jet axis towards the core. It can be seen clearly that a sharp filament exists along the base of the western lobe, while the eastern lobe shows two relatively faint scythe-like structures. These structures, labelled `scythe1' and `scythe2' in the bottom panel (see inset), as well as the filament, are strikingly sharp and seem to flow transverse to the jet axis (see Figure \ref{fig:3c320_cband} to see the detected jet at 6000 MHz). We discuss the nature of these features further in Section \ref{sect:environment}. 

3C\,320 has been scarcely studied in the past, prohibiting a detailed multi-wavelength analysis based on the literature. The only high resolution (sub-arcsec) study of this source was presented by \cite{maha20} with VLA observations at 1500\,MHz and 6000\,MHz, but the scythes and filament are only seen with the 144\,MHz data. We also detect diffuse extension of material just east of the eastern lobe, where the lobe boundary is expected to exist. This would be expected under a model in which the jet termination occurs on the front or back boundary of the lobe in projection, as suggested for wide-angle tail radio galaxies \citep[][]{hard04}. We note that this extension is seen at 1500\,MHz from the data presented by \cite{maha20}, but not at 6000\,MHz, consistent with real  steep-spectrum emission.

\subsection{Environment}
\label{sect:environment}

\subsubsection{3C34}
In the bottom panel of Figure \ref{fig:3c34} we present the LOFAR image overlaid with HST contours at $\lambda=8620\AA$ from \cite{best97}. The host galaxy has spheroidal morphology and is surrounded by diffuse optical emission (due to a superposition of cluster galaxies in the line of sight, as suggested by \citealt{mcca95}). There is also another large spheroidal cluster member to the north. However, there are no signs such as tidal features in the optical images of an interaction pre- or post-merger, suggesting that lobe-merger interactions are not responsible for the morphology of the lobes, or for the stripping of lobe material to form the diffuse tails in the north-south direction.

Noticeably the large central filament is curved, in projection, around the host galaxy, presumably reflecting the galaxy halo which has a higher density than the surrounding ICM. This would indicate the origin of the filament is related to the high density regions in the ambient medium -- there is either a density jump in the ambient medium encountered by the back-flowing plasma, causing a shock to be developed, and the back-flowing aged particles are re-energized in this shock front, and/or magnetic fields are particularly ordered and enhanced  around the dense region. This is discussed further, through evidence of a spectral index flattening at the location of the filament, in Section~\ref{sect:particle_acceleration}. 

`Object a' is a thin strip seen in optical and far-infrared images, interpreted as being a region of a jet-cloud interaction driving star-formation in a galaxy \citep[][]{best97}. However, as we discuss in Section \ref{sect:discussion_particleacceleration} below, this is not in line with the detected jet axis (see Figure \ref{fig:3c34_cband}), which aligns with the southern hotspot in the western lobe with the core and the jet knots in the eastern lobe. Moreover, there is a vertical bridge (on scales of a few tens of kpc) that seems to perpendicularly cross `Object a' and bridges the radio limb brightening either side (north-south) of `Object a'. This region, termed as a depolarization silhouette by \cite{john95}, exhibits strong depolarization between 1 GHz and 5 GHz compared to its surroundings -- defining the depolarization $DP^{21}_{6}$ as the ratio of the percentage polarization between 21\,cm and 6\,cm, \cite{john95} find $DP^{21}_{6}\sim0.1$ directly at the location of `Object a' and the vertical bridge and $DP^{21}_{6}\sim0.5$ in the immediate surroundings. This implies that the depolarizing medium, likely to be a high thermal gas density region, has suitable conditions for filament formation. A high thermal gas density would also explain the limb-brightening either side of the vertical bridge, if the back-flowing plasma naturally avoids high density regions, consistent with our picture for the central filament.


\subsubsection{3C320}
In the bottom panel of Figure \ref{fig:3c320} we display the 144\,MHz LOFAR image with 0.5--5\,keV \textit{Chandra} contours overlaid, taken from \cite{maha20}. A thermal ICM shock driven by the radio lobes \citep[][]{vags19,maha20} is clearly seen to surround the radio source. Interestingly, a high surface-brightness bar of emission at the centre of the ICM (marked as `ICM bar') aligns well with `scythe2' and the central filament seen across the base of the lobes, as marked in Figure~\ref{fig:3c320}. It is possible that back-flowing plasma is disrupted by the presence of this high density region, or that ordered magnetic fields play a role in enhancing the radio emission, as might be the cases for the origin of the features seen in 3C\,34. We discuss the implications for the spatial coincidence between filamentation and the dense regions at the centre of the ICM in Section \ref{sect:discussion_filaments}. 

\subsection{Spectral structure \& particle acceleration}
\label{sect:particle_acceleration}

\begin{figure*}
    \centering
    \hspace{-0cm}\includegraphics[scale=1.5,trim={3.5cm 0.4cm 3.5cm 0cm},clip]{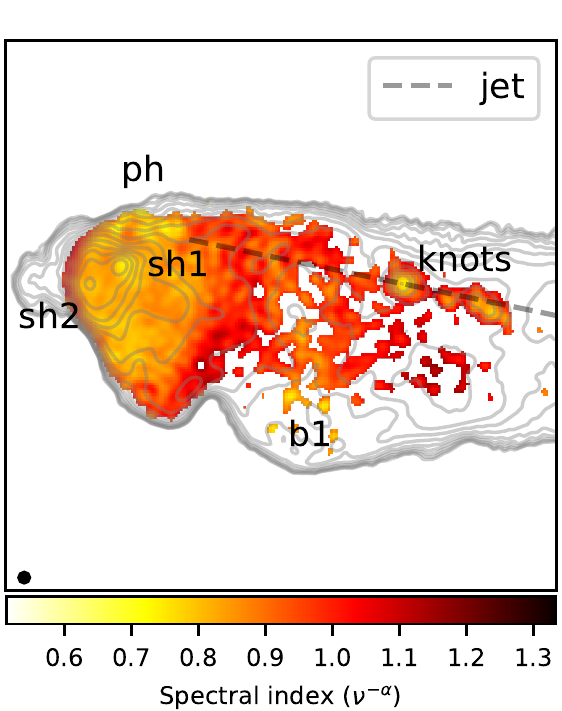}\hspace{0.5cm}
    \includegraphics[scale=1.5,trim={3.5cm 0.4cm 3.5cm 0cm},clip]{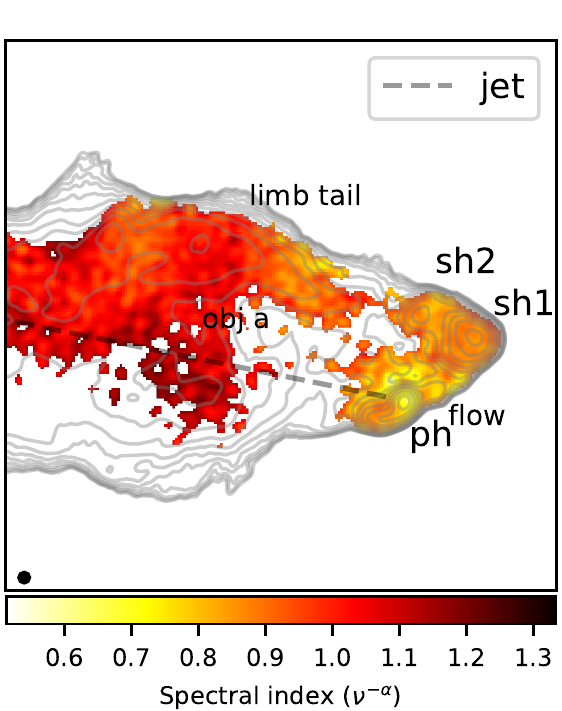}
    \caption{Spectral index maps of the eastern (left) and western (right) hotspot structures of 3C34. Maps are of two-point spectral indices between 144\,MHz and 5000\,MHz, convolved to a circular 0.38 arcsec beam (black circle on the lower left). Total intensity contours of the 144\,MHz data are overlaid, starting at 20$\sigma$ (to emphasize the brighter and more compact regions), multiplied by 1.5 to the power of integers from 1 to 20. }
    \label{fig:3c34_ehotspot}
\end{figure*}


\subsubsection{3C34}
3C\,34 clearly has multiple compact features in both lobes. While the exact definitions of hotspots and jet knots are debated, we use the traditional terminology: hotspots are distinct compact structures found at the edges of an FR-II source associated with jet termination, whereas jet knots are compact features as part of a well-collimated jet. Where multiple hotspots are detected in a single lobe, as is particularly the case for 3C\,34, we adopt the term `hotspot complex' used by \cite{leah97} to refer to a group of hotspots at the end of a given lobe that is not associated with a jet knot. Following \cite{lain89}, we define the `primary hotspot' as the most compact hotspot, and all other compact features in the hotspot complex are `secondary hotspots'.  

In Figure \ref{fig:3c34_ehotspot}, we display two-point spectral index maps between 144\,MHz and 5000\,MHz of the eastern and western lobes, at an angular resolution of 0.38 arcsec. In the eastern lobe (left panel) there are many visible compact features evident through the LOFAR total intensity contours (grey); the jet knots are resolved into four components and least two bright hotspots at the eastern edge. We note that, as suggested by \cite{john95} and \cite{mull06}, the jet axis is most likely that which aligns the jet knots in the eastern lobe with the jet that is well detected in the western lobe (as marked in the figure). The spectral index map reveals that there is a new region containing flat spectrum components -- the expected jet termination point contains a region with $\sim3$ point sources (yellow in colour) with the flattest spectral index in the lobes with $\alpha_{5000}^{150}\approx0.65\pm0.10$. Since these are the most compact and flat-spectrum components residing in the hotspot complex, and since they lie at the expected jet termination point, we associate this region with that containing the primary hotspot (labelled `ph'), although clearly this region itself can be called a `primary hotspot complex'. 

The secondary, brighter and larger hotspot just south of the primary hotspot region (`sh1') has $\alpha_{5000}^{150}\approx0.75\pm0.06$, and the other secondary hotspot to the south-east of this (`sh2') has $\alpha_{5000}^{150}\approx0.85\pm0.06$. A strikingly similar effect is seen for the western lobe on the right panel of Figure \ref{fig:3c34_ehotspot}, with the most southern and primary hotspot (`ph') having $\alpha_{5000}^{150}\sim0.72\pm0.06$ and the secondary larger hotspot at the leading western edge (`sh1') having $\alpha_{5000}^{150}\sim0.91\pm0.06$. As in the eastern lobe, `sh2' in the eastern lobe is slightly less compact and further downstream from `sh1'. There is a region of flat spectrum emission between the `ph' and `sh1', which we have called the `flow' region, with $\alpha_{5000}^{150}\sim0.69\pm0.12$, which is flatter than the secondary hotspot and similar in spectral indices with `ph'. The total intensity contours show that emission extends north-west of `ph' toward the secondary hotspots, implying a flow of plasma. In both lobes, it is clear that if the newly found primary hotspots are undergoing particle acceleration, they are not restricted to one location in their primary hotspot complexes, and that the spectral indices of secondary hotspots steepen with distance from the primary hotspot. This has implications for the dynamics of particle acceleration, jet-lobe interactions and the origin of the secondary hotspots, which we discuss in Section \ref{sect:discussion_particleacceleration}. 
\begin{figure}
    \centering
    \includegraphics[scale=1.0,trim={1.9cm 0.3cm 1.7cm 0.2cm},clip]{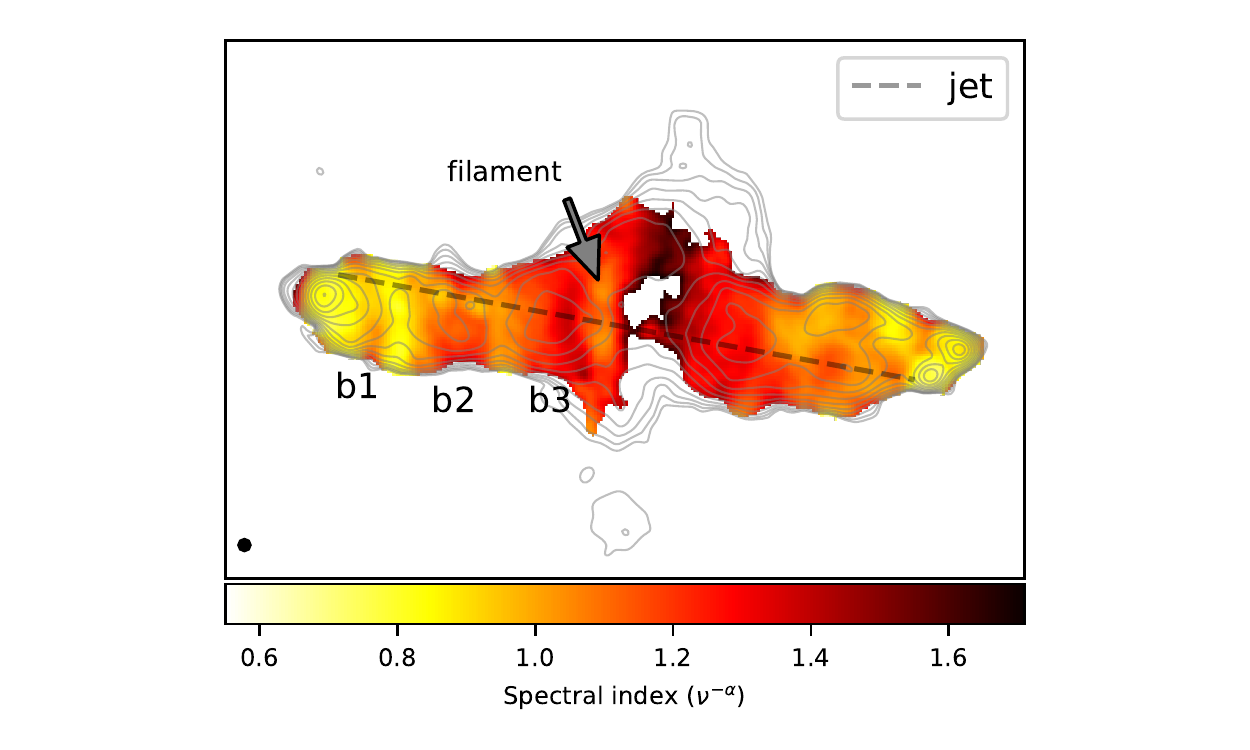}
    \includegraphics[width=9cm,trim={0.5cm 1.3cm 0.5cm 0.7cm},clip]{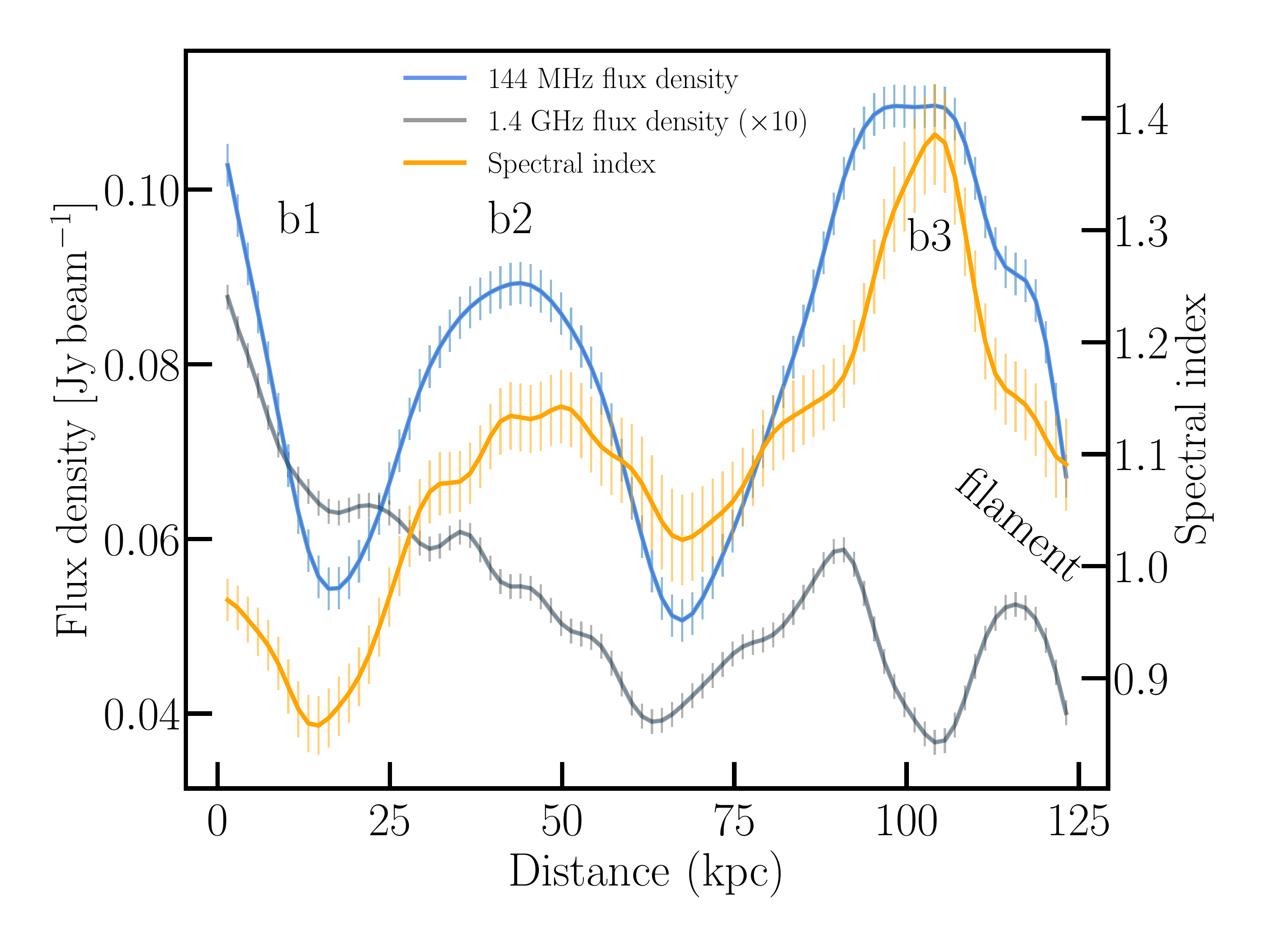}
    \caption{Top: Two-point spectral index map between 144 MHz and 1400 MHz for 3C\,34, for the regions detected above 5$\sigma$, with a common beam size of 1.2 arcsec. Contours are of the 144 MHz LOFAR image at 1.2 arcsec resolution. Bottom: Radial line profile along the major axis of the eastern lobe of 3C\,34 in total intensity at 144\,MHz (blue), 1400\,MHz (grey) and spectral index between 144\,MHz and 1400\,MHz (orange). Error bars describe the measurement uncertainties. The locations of the banding regions and the central filament are annotated.}
    \label{fig:3c34_spix}
\end{figure}

Another curious aspect of 3C\,34 is the periodic surface brightness banding. As can be seen in the LOFAR image on the top panel of Figure \ref{fig:3c34}, the eastern lobe shows periodic patches of surface brightness increases and depressions from the hotspots to the core. In the top panel of Figure \ref{fig:3c34_spix}, we display the two-point spectral index map at 1.2 arcsec resolution, using the 144\,MHz and 1400\,MHz data, overlaid with the 144\,MHz total intensity contours at the same resolution. The banding pattern (labelled `b1', `b2' and `b3') is most prominently seen in the 144\,MHz data, and it can be seen that the spectral indices flatten in the surface brightness depressions (see grey LOFAR contours) between the banding. In the bottom panel of Figure \ref{fig:3c34_spix} we plot flux densities at 144\,MHz and at 1400\,MHz (multiplied by 10 for presentation purposes) and spectral index across a 1-D radial line profile from b1 to the central filament (perpendicular to the centre of the minor axis of the lobe). We include measurement uncertainties only, using error propagation for the spectral indices, to reflect more appropriate uncertainties on pixel to pixel variations of diffuse lower surface brightness emission. It can clearly be seen that the low frequency (blue line) banding represents regions of higher surface brightness and steeper spectral index than the regions in between the bands, which show a clear spectral flattening; this is inconsistent with gradual spectral ageing from the hotspots toward the core. The total intensity banding at b1 and b2 is less clear at 1400\,MHz, but the depression between b2 and b3 and the steep decline at b3 is clearly consistent with banding. The spectral flattening to the west of b3 is particularly interesting, as the flattening is precisely at the location of the central filament, as marked in Figure \ref{fig:3c34_spix}. Further downstream, towards the wing just north-west of the filament, the spectral index steepens again to $\alpha_{1400}^{150}\geqslant$1.5. To our knowledge such patterns have not been observed in total intensity for radio galaxies, although banded rotation measure patterns in radio galaxies, associated with magnetic field compression of an external magnetized medium has been reported by \cite{guid11}. Meanwhile, the spectral flattening of the central filament may indicate enhanced magnetic fields within the source, or a re-energisation of back-flowing material. We discuss this further in Section \ref{sect:discussion_filaments}.

\subsubsection{3C320}
\begin{figure*}
    \centering
    \includegraphics[scale=1.6,trim={1cm 0.4cm 0.9cm 0cm},clip]{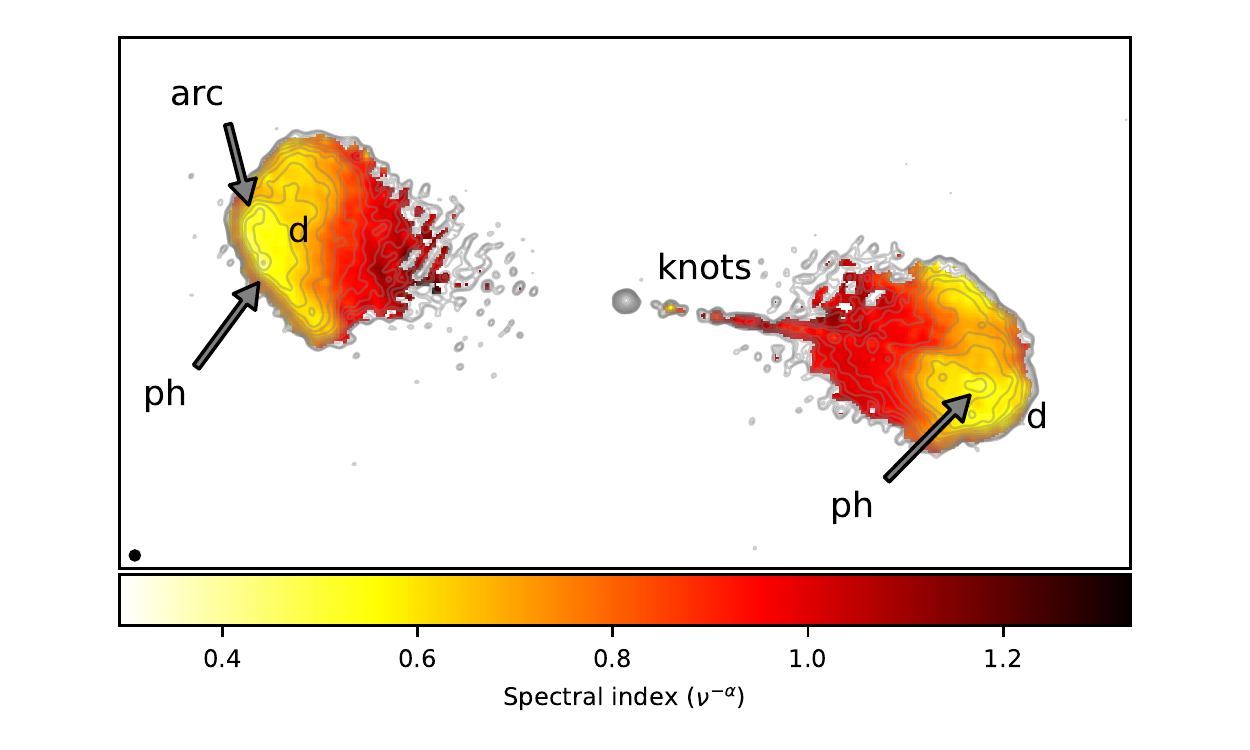}\\
    \caption{Four-point spectral index map between 150 MHz and 6000 MHz for 3C\,320, for the regions detected above 5$\sigma$, with all three maps convolved to a common beam size of 0.2 arcsec (black circle). Contours are of the VLA 6000\,MHz total intensity emission at the same resolution. The locations of high energy emission discussed in the text are annotated. The jet axis is not marked due to a lack of a clear counter-jet detection, but is clearly seen on the jetted side.}
    \label{fig:3c320_spix}
\end{figure*}
3C320 is known to have a complex hotspot behaviour in the eastern lobe when viewed at sub-arcsecond resolution \citep{maha20}, while the western lobe shows only one compact flat-spectrum region, consistent with the findings of \cite{brid94} that more complex hotspot structures are favoured on the counter-jet side \footnote{It should be noted that a jet is also detected on the western side, albeit much fainter than the eastern side, and hence the jets are orientated close to the plane of the sky, as per the case of 3C34.}.

In Figure~\ref{fig:3c320_spix} we display the four-point spectral index map of 3C\,320 between 144\,MHz and 6000\,MHz. For purposes of better displaying the distinction between flat and steep spectral structure in the lobes, we have saturated the colour scale for $\alpha\leqslant0.3$ where the radio core, which is undetected or is dimmer than the surrounding lobe material in the 144\,MHz data, shows an inverted spectrum. The jetted, western lobe shows a single compact component in total intensity with a spectral index ($\alpha=0.50\pm0.08$), which we associate with the primary hotspot (labelled `ph' in Figure \ref{fig:3c320_spix}). This would require the jet to bend slightly to the south in order to terminate at `ph'. The most interesting aspect is that flat-spectrum emission is seen ahead of the primary hotspot in the form of diffuse components, labelled `d'. Recessed hotspots with lobe material further upstream are common, representing the case where the jet interacts with the front or back boundary of the lobe in projection. The upstream diffuse material (`d') also has flat spectral indices ($0.52\leqslant\alpha\leqslant0.57$) but with values consistent with the suggestion that the compact region ph is the primary hotspot. There is a region at the northern edge of the western lobe containing flat spectral indices with $\alpha=0.56\pm0.23$. This feature is similar to the limb tail seen in the northern edge of the western lobe of 3C\,34, but it is not associated with any obvious distinct structure in any of the total intensity maps of 3C\,320.

The eastern lobe also shows a single compact flat spectrum component with $\alpha=0.46\pm0.08$, which we associate with the primary hotspot label `ph'. This hotspot seems to be associated with an arc (labelled `arc') of similarly flat-spectrum emission, akin to those found for the northern hotspot of 3C\,33 \citep[][]{rudn90} and the southern hotspot of 3C\,445 \citep[][]{prie02}. We again see a diffuse region of emission with $\alpha=0.55\pm0.10$, although this is behind the primary hotspot, unlike the case for the western lobe. The fact that the eastern hotspot is not recessed is consistent with it being on the counter-jet side, as is commonly the case \citep[][]{brid94}. The origin and nature of the diffuse flat-spectrum regions in both lobes is unclear, but nevertheless give hints toward distributed particle acceleration, commonly associated with FR-I jets. We discuss the implications of these results in Section \ref{sect:discussion_particleacceleration}.

Unfortunately, the scythes and filament are not detected at higher frequency, prohibiting a detailed spectral analysis. However, the non-detection of the scythes and filament at 1500\,MHz suggests that the spectrum must be steep having $\alpha^{144}_{1500} > 1.2$ in the brighter regions closer to the host galaxy, and $\alpha^{144}_{1500} > 1$ in fainter regions farther away.\footnote{Here we have assumed a $2\sigma$ non-detection limit at 1.52\,GHz for an RMS noise of $\rm 50\,\upmu Jy\,beam^{-1}$.} This clearly indicates that the synchrotron emitting plasma in the filament are composed of relatively old cosmic ray electrons (CREs), rather than newly accelerated CREs of the type expected to exist at hotspots. Nevertheless, our new LOFAR image shows evidence for clear interaction between the oldest plasma (> 25 Myr, based on the spectral ageing analysis of higher frequency data by \citealt{maha20}) and its surrounding environment, other than the presence of large scale X-ray cavities coincident with the lobes \citep{vags19}.

Using the lower limit on the spectral index of 1 in the filament, and assuming a cylindrical geometry of the filaments having path length ($l$) similar to the projected-width on the plane of the sky ($l = 3$\,kpc for a width of 0.6\,arcsec at the redshift of 3C\,320), we estimated the equipartition magnetic field strength $\lesssim 15\,\upmu$G by using the revised equipartition formula \citep{beckkrause05}.\footnote{Here we have used the ratio of the number densities of relativistic protons to electrons $K_0 = 0$ and a volume filling factor of 1.} This provides a lower limit on the synchrotron+inverse-Compton cooling lifetime $\gtrsim 50\,\text{Myr}$ at 144\,MHz in the filament, a factor of two from the lower limit lifetime of $25$ Myr found from higher frequency data \citep[][]{maha20}.

\section{Discussion}
\label{sect:discussion}
\subsection{Models for particle acceleration and jet termination}
\label{sect:discussion_particleacceleration}
The addition of low frequency (144\,MHz) data has unveiled new hotspot structures and other areas of flat-spectrum emission in 3C\,34 and 3C\,320, while the sub-arcsecond resolution maps have helped to distinguish between compact and more diffuse structures. We note that the distinction between sites of particle acceleration and ageing is best given by evidence for optical and X-ray synchrotron emission in hotspot complexes, while in the radio regime the highest available frequencies ensures sensitivity to only the most recently accelerated particles. Nevertheless, with our spectral index maps and with the overall morphology seen at sub-arcsec resolution between 144\,MHz and 6\,GHz we can make qualitative statements about jet termination models for 3C\,34 and 3C\,320. 

Flat-spectrum emission at the jet termination point is not restricted to one location -- distributed particle acceleration is suggested to occur in the hotspot complexes of radio galaxies, particularly in secondary hotspots \citep[e.g.][]{hard07}, and also on physically larger scales in the jets of FR-I sources \citep[e.g.][]{evan05}. Using spectral index maps we discover new regions (that are not clearly visible in total intensity but are detected at >$10\sigma$) that are directly associated with the jet termination in both lobes of 3C\,34 (`ph' or primary hotspot complex; Figure \ref{fig:3c34_ehotspot}), which are multi-component and have sizes of $\sim2\rm\,kpc$. The secondary hotspots, having sizes larger than 4\,kpc, have a steepening with distance from the primary hotspot complex. On the other hand, 3C\,320 displays only one compact component in both lobes, and flat-spectrum diffuse and arc-like components (Figure \ref{fig:3c320_spix}) exist in the vicinity of the primary hotspot. Competing theories exist for the formation of secondary hotspots: models that propose that all hotspots are connected to the energy supply include the `splatter-spot' model \citep[][]{will85}, in which jet material flows out of the primary hotspot and terminates again in the locations of secondary hotspots. Other models suggest that the jet end-point moves along the lobe minor axis, such as the `dentist-drill' \citep[][]{sche82} and precessing jet \citep[][]{cox91} models. The major difference in physical processes between these models is that, in the case of the dentist-drill or precession models, secondary hotspots are remnants left behind by the motion of the jet where particle acceleration should cease, causing a steepening of their spectra at high frequencies. On the other hand, if secondary hotspots continue to be fed by the energy supply (i.e. the splatter-spot model), particle acceleration will continue in them, and we expect flat-spectrum emission at high frequencies. Note that jet precession is generally invoked as a prerequisite for the dentist-drill model where the tip of the jet asymmetrically changes direction, but jets are also known to abruptly change direction upon entering the lobe \citep[e.g. 3C\,175][]{kell88}. For our purposes, we reference both models to suggest the same jet termination model, as we only distinguish between jet models describing hotspots that are all connected with the energy supply and those that are not. 

3C\,34 displays secondary hotspots that have more aged plasma with distance from the primary hotspot(s) -- this provides strong evidence for the dentist-drill or precessing jet models acting in this system. This was suggested by \cite{best97}, through the absence, with their data, of a clear termination point of the jet head in the eastern lobe. Furthermore, this would also explain the jet-induced optical strip `Object a', which does not lie along the current jet direction but may have been formed from the jet from a previous precession angle. If the previous state of the jet was such that it connected the core with `Object a' and `sh1' in a straight line (see right panel of Figure \ref{fig:3c34_ehotspot}), then this would require the jet to have precessed by 6$^{\circ}$ (in projection) in the clockwise direction to its current position (marked as `jet' in Figure \ref{fig:3c34_ehotspot}), consistent with precession angles inferred from hotspot separations in 3C sources \citep[][]{best95}. The gradual steepening of spectral index from the primary hotspots, particularly seen in the `flow' region in the western lobe, gives a consistent picture. The splatter-spot model cannot be ruled out, but would require extreme deflection of the jet flow, as it would have to bend $\sim90^{\circ}$ to the south from `ph' to `sh1' in the eastern lobe. Moreover, the splatter-spot model cannot be reconciled with our spectral index maps (since it would require the secondary hotspots to have similarly flat spectral indices as for the primary hotspots).

3C\,320 does not display clear secondary hotspots, but rather diffuse flat-spectrum regions in front (western lobe) and behind (eastern lobe) the primary hotspot, as seen in Figure \ref{fig:3c320_spix} (marked as `d'). Diffuse and distributed particle acceleration, as is evident through X-ray synchrotron emission, is usually seen in FR-I jets, but has also been reported in some FR-II sources (e.g. 3C\,390.3; \citealt{hard07} and 3C\,33; \citealt{kraf07_3c33}) where the radio emission is often either weak or not correlated with the X-ray morphology. With the current data we cannot provide strong evidence for particle acceleration in these regions due to a lack of high resolution X-ray data, but nevertheless the spectral information favour the splatter-spot model in which flat spectrum emission is evident in the primary hotspots and the diffuse regions. While the dentist-drill or precession models would be consistent with the radio morphology (due to the presence of diffuse flat-spectrum regions and the arc in the eastern lobe), the spectral characteristics in the hotspot complexes would require steepening. \cite{rudn90} discussed these models to describe the northern hotspot complex of 3C\,33, also containing a single hotspot and an arc connecting the hotspot on the outer edge of the lobe. Without spectral information, they could not discriminate between the splatter-spot and the dentist-drill models. In any case, we cannot rule out ageing and structures that are not connected with particle acceleration, and therefore discrimination against any particular model, if the break frequency occurs beyond 6000\,MHz. More robust spectral information in this regard would require higher frequencies -- observations of primary hotspots show that a spectral cut-off tends to be present at millimetre wavelengths ($\sim30$\,GHz).  

We conclude that, due to the differences in morphology and spectral structure between 3C\,34 and 3C\,320, powerful sources in cluster environments can require different jet termination models to account for their energetics. We confirm, along with many other studies, that the working surface at the head of the jet is not static, but dynamic. The hotspots in a given lobe can vary in size, shape, position and spectral indices as a function of time, and has clearly been demonstrated with numerical simulations for many years \citep[e.g.][]{burn91}. The most favourable models, using the available data, are the dentist-drill or precession models and the splatter-spot model, for 3C\,34 and 3C\,320 respectively. If this is true, further work must be undertaken to understand why cluster-environment sources differ in terms of their jet-lobe interactions. 
\subsection{Nature and origin of the oldest lobe plasma: filaments, scythes and banding}
\label{sect:discussion_filaments}
At 144\,MHz and at 0.3 arcsec resolution, we have presented the first high fidelity images of the radio sources 3C\,34 and 3C\,320 in order to understand the structure of the oldest radio galaxy plasma in rich environments. We have detected lobe banding in 3C\,34, scythe-like filamentation in 3C\,320 and large-scale (> 40 kpc) filaments near the base of the lobes in both sources, all of which have steep spectra indicating their association with the oldest plasma.

For both sources, the nature of the central filaments is clearer through the interaction with their environments. For 3C\,34 the filament bends around the host galaxy, or the halo of the host galaxy, evident in the bottom panel of Figure \ref{fig:3c34}. For 3C\,320 the filament (along with `scythe2'; Figure \ref{fig:3c320} bottom panel) lies at a surface brightness jump toward the central dense bar in the ICM. For both sources, the remarkable alignment of central filamentation with the dense material in the surrounding medium (as also recently seen in NGC\,507; \citealt{brie22}) implies a likely physical association between the non-thermal radio plasma and the surrounding thermal plasma. Similar situations have been observed recently for the cases of Nest200047 \citep[][]{brie21} and Abell\,2626 \citep[][]{igne20}, where in the former study the combination of buoyancy forces and cluster motions (e.g. gas `sloshing') is suggested to drive material away from the lobes and produce filamentation. The strikingly large and collimated nature of filaments at the bases of the lobes in 3C\,34 and 3C\,320 cannot solely be explained by this process since the lobe material immediately surrounding the filaments would also be affected, in a morphological sense, unlike in the aforementioned cases where large-scale lobes do not surround the filaments.

Large-scale ordered and/or compressed magnetic fields in the ICM in the shape of the filaments can also contribute to their increased surface brightnesses relative to their surroundings -- magnetohydrodynamic simulations have suggested that filaments trace enhancements in the magnetic field distribution in radio galaxies \citep[][]{huar11,hard13}. This would be qualitatively consistent with the implication from Figure \ref{fig:3c34_spix}, showing a spectral flattening of aged plasma by $\sim20\%$ (the peak spectral index of b3 is $\sim1.4$ while that for the filament is $\sim1.15$) at the location of the filament seen at 144\,MHz. This flattening, which deviates from simple models of spectral ageing with a smooth steepening from hotspot to core, may also indicate that the filament contains re-energized particles -- while the spectral index of the filament is significantly steeper than what is expected for Fermi-I particle acceleration, this spectral behaviour is analogous to the process of `gentle re-energisation' of particles introduced by \cite{dega17}. In this case, downstream from the radio lobes, tails at very low frequencies (< 100\,MHz) can be observed, with only a slight flattening of the spectral index between the base of the lobes and the tails, and such mild flattening is associated with inefficient particle acceleration that is \textit{just} balanced by radiative cooling. If such a process is occurring in 3C\,34 and 3C\,320, this has major implications for radio plasma physics: with such a mild flattening, an additional particle acceleration process other than diffuse shock acceleration and Fermi-I (thought to occur in hotspots) may be present that is particular to cluster and high density environments surrounding radio galaxies. Moreover, this scenario alludes to the existence of long-lived high energy cosmic rays that must exist in the ICM as a seed population that are later re-accelerated by cluster processes to form haloes and relics \citep[see review by][]{vanw19}. Clearly, the oldest plasma extends outwards from the filaments in the north-south direction, thereby distributing plasma into the peripheries of the ICM. Even without particle re-acceleration, it is clear that magnetic field enhancement plays a role in developing large-scale filaments. Deeper data at sub-arcsec resolutions over a large frequency range that can detect these filaments are needed to firmly establish these results. Nevertheless, our results show a clear association and interaction between large-scale filamentation and the densest regions of the ICM.

The inner parts of the low surface brightness tails were previously identified as wings from a large-scale backflow \citep[][]{john95}. We see that these diffuse tails are almost perpendicular to the jet axis. X-shaped radio galaxies, where the lobe structure is seen to extend perpendicular to the jet axis \citep[e.g. PKS 2014-55][]{cott20} have been observed for many years, but the origin of the perpendicular wings is still a debate -- rapid changes in the black hole spin direction \citep[][]{robe15}, multiple pairs of jets \citep[][]{brun20} and jet deflection by the hot halo of the host galaxy \citep[][]{cott20} have all been suggested to explain such morphology. We prefer the explanation given by \cite{cape02}, who found a striking correlation between the axes of wings and those of host galaxies of radio sources -- the wings are preferentially directed along the minor axis of their elliptical host galaxies, and there is a preference towards high projected ellipticity. This naturally allows the backflow to expand laterally due to the steeper pressure gradient in the ICM perpendicular to the jet axis, and has been modelled with numerical simulations \citep[][]{leah85,cape02,giri22}. The HST image in Figure \ref{fig:3c34} does not conclusively show an elliptical morphology but rather a spheroidal morphology for the host, which \cite{mcca95} attribute to a superposition of cluster members in the line of sight. Given further that the orientation of the host could be such that its major axis is inclined to the line of sight, we cannot conclusively reject the model of \cite{cape02}. 

The simulations of \cite{giri22} also predict that the turbulent nature of the plasma in wings or diffuse tails will induce in situ re-energisation of particles through local shocks, causing the wing material to survive over longer time scales than what would be expected under a model of only radiative and adiabatic losses. More sensitive observations at high resolution at frequencies greater than 1\,GHz that would detect the tails will allow for a highly resolved spectral analysis of the diffuse tails to confirm this scenario, for the case of 3C\,34. It is also unclear whether the tails are physically associated with the central filament (or if the scythes and filament are associated with the lateral lobe expansion in 3C\,320), but if the filament contains enhanced magnetic fields, this would provide a mechanism for material to be transported along the filament and toward the ICM outskirts -- as would be needed to support theories for the formation of diffuse cluster radio sources.

The `edge' tail seen on the south side of the eastern lobe of 3C\,34 (see Figure \ref{fig:3c34}) is a newly seen structure and is seen to extend downstream towards the base of the lobes. A consistent explanation with the aforementioned analysis that filaments and tails are shaped by the gas distribution of the ambient medium can be made if the edge tail is produced by lobe-environment interactions -- Figure \ref{fig:3c34} (bottom panel, see annotation) shows that the tail is highest in surface brightness at a location where there is a `dent' at the lobe southern edge. The HST image shows no galaxy or source in this location, and any interaction would most likely arise from clumps of hot gas in the ICM. To our knowledge this is the first example of filamentation along the inner edge of a radio galaxy lobe and deeper observations, particularly in the X-ray, are required to give a view of the ICM gas distribution.

The nature of the lobe banding in the eastern lobe of 3C\,34 (Figure \ref{fig:3c34_spix}) is particularly difficult to understand with the current data. This is the first case in radio galaxy lobes, to our knowledge, where there are periodic large (few tens of kpc) diffuse patches of high surface brightness which are steep-spectrum, relative to their surroundings along the jet axis. \cite{guid11} reported banded Rotation Measure ($RM$) structure in the lobes of a small sample of radio galaxies, suggesting that the corresponding Faraday rotation occurs due to the lobes expanding into surrounding magnetized plasma, compressing and enhancing the magnetic fields in the immediate surroundings. Radio lobes are known to be in contact or to interact with their environment -- the presence of X-ray cavities seen in a number of objects situated in clusters \citep[e.g.][]{mcnamara2007} suggests radio galaxy lobes in general can displace ambient thermal gas. In the case of 3C\,34 the banded patches have a steeper spectrum than their surroundings (see Figure \ref{fig:3c34_spix}), requiring lower magnetic field strengths if the intrinsic electron spectrum is curved in this region, inconsistent with the field compression scenario. However, the regions between the bands do have flatter spectral indices (see `b1' in Figure \ref{fig:3c34_ehotspot}), which may indicate regions of magnetic field compression, but such narrow linear features expanding into the magnetized ICM quicker than the surrounding lobe material is implausible. 

The radio galaxy Hercules A has been observed to show spectral differences between prominent ring structures and the surrounding lobe material in the western lobe, suggested as being associated with cyclic activity \citep{timm22}, and such sharp spectral boundaries have been used to suggest cyclic jet activity in other sources (e.g. 3C\,388; \citealt{roet94}). 3C\,34 on the other hand does not show any prominent and sharp features that mark the boundary of the bands of diffuse steep-spectrum material. A more convincing explanation for the banding is that the backflow is periodic (rather than the jet activity) driven by jet precession, which we have associated to the jets of 3C\,34 due to the spectral properties of the hotspots. For a precessing jet, particularly if the precession properties change with time (simulations suggest wider precession cone opening angles lead to slower growth, e.g. \citealt{hort20}), the forward expansion rate is likely to be periodic. Therefore, the backflow is likely to be periodic, leading to a pattern of long and short-lived electron populations across the lobes toward the core. By calibrating radio source evolution models with radio observations, \cite{turn18_3cmodels} determine the age of 3C\,34 to be 25.1\,Myr. If there are indeed three bands of distinct electron populations driven solely by jet precession, then this constrains a potential precession period of 8\,Myr, consistent with precession periods of 3C sources (of the order of 1–10 Myr) based on the analysis by \cite{krau19}. Nevertheless, we can conclude that one of either standard dynamics or standard shock acceleration mechanisms does not apply in the formation of the banding. Further studies of other cluster-centre sources at low frequencies will shed light on the ubiquity of such structures and will help to determine empirical trends that challenge standard models of radio galaxy evolution. 

\section{Conclusions}
\label{sect:conclusions}

Our high resolution imaging of 3C\,34 and 3C\,320 at 144\,MHz has revealed the complexities associated with both the highest and lowest energy radio emission observed in the lobe and hotspot plasma. In summary, our conclusions are:
\begin{itemize}
    \item Large-scale filamentation spanning the width of the lobes of powerful radio galaxies may be a common phenomenon in cluster environments. The precise shape and location of the filaments is associated with the density profile of the ambient environment.
    \item For 3C\,34 we find a spectral flattening at the location of the filament, suggesting that enhanced magnetic fields are responsible for its formation. The resulting particle populations, that may be re-energized, continue to be driven  perpendicular to the jet axis to form low surface brightness tails that can mix with the ICM, and provides support to radio galaxy lobes providing the seed particles for the existence of diffuse cluster sources.
    \item We report the first case of large-scale periodic banding in total intensity and spectral index in the lobes of radio galaxies, evident in the eastern lobe of 3C\,34. The origin of these structures is unknown, but may be a result of jet precession.
    \item The dynamic hotspot structure in the lobes reveal that the jet termination point is fragmented into multiple closely situated regions of very compact regions of similar spectral indices, but also diffuse regions of flat spectral index in 3C\,320. We argue that the dentist-drill or precessing models and the splatter-spot model best describe the jet behaviours in 3C\,34 and 3C\,320, respectively. 
\end{itemize}

\section*{Acknowledgements}
We thank the anonymous referee for helpful insights that has improved the paper. We also thank Andrea Botteon and Chris Riseley for helpful comments that improved the paper. 

MJH acknowledges support from the UK STFC [ST/V000624/1]. This work was supported by the Medical Research Council [MR/T042842/1]. RJvW acknowledges support from the ERC Starting Grant ClusterWeb 804208. 

This research made use of the University of Hertfordshire
high-performance computing facility and the LOFAR-UK computing
facility located at the University of Hertfordshire (\url{https://uhhpc.herts.ac.uk}) and supported by
STFC [ST/P000096/1].

LOFAR is the Low Frequency Array, designed and constructed by ASTRON. It has observing, data processing, and data storage facilities in several countries, which are owned by various parties (each with their own funding sources), and which are collectively operated by the ILT foundation under a joint scientific policy. The ILT resources have benefited from the following recent major funding sources: CNRS-INSU, Observatoire de Paris and Université d'Orléans, France; BMBF, MIWF-NRW, MPG, Germany; Science Foundation Ireland (SFI), Department of Business, Enterprise and Innovation (DBEI), Ireland; NWO, The Netherlands; The Science and Technology Facilities Council, UK; Ministry of Science and Higher Education, Poland; The Istituto Nazionale di Astrofisica (INAF), Italy.

The National Radio Astronomy Observatory is a facility of the National Science Foundation operated under cooperative agreement by Associated Universities, Inc.

This research made use of APLpy, an open-source plotting package for Python \citep{robi12}.
\section*{Data Availability}
The data underlying this article will be shared on reasonable request to the corresponding author.



\bibliographystyle{mnras}
\bibliography{references} 






\bsp	
\label{lastpage}
\end{document}